\documentclass[journal,onecolumn]{IEEEtran}
\usepackage{epsfig,color,amsmath,cite}

\IEEEoverridecommandlockouts
\usepackage{wrapfig}
\usepackage{setspace}  
\usepackage{xcolor}
\usepackage{epstopdf}
\usepackage{amssymb}
\usepackage{url}
\usepackage{mathtools, nccmath}
\usepackage{enumitem}
\usepackage{multirow}
\usepackage{makecell}
\usepackage{amsmath}
\usepackage{stackrel}
\usepackage{booktabs}
\usepackage[flushleft]{threeparttable}
\usepackage{makecell}

\usepackage{subfig}
\usepackage{graphicx}
\usepackage{lscape}

\usepackage{comment}
\usepackage{chngpage}
\usepackage{pifont}

\usepackage[linesnumbered,lined,boxed,commentsnumbered,ruled,longend]{algorithm2e}
\makeatletter

\SetKwProg{Init}{Initialization}{}{Proceed}
\SetKwProg{St}{Start}{}{Proceed}
\makeatother

\setlist[itemize]{leftmargin=*}

\DeclareMathOperator{\rank}{rank}

\DeclareMathOperator*{\minimize}{minimize}

\DeclareMathOperator{\subjectto}{subject\ to}

\makeatother

\newcommand{\mat}[1]{\boldsymbol{#1}}
\renewcommand{\vec}[1]{\boldsymbol{\mathrm{#1}}}

\providecommand{\mA}{\ensuremath{\mat{A}}}
\providecommand{\mB}{\ensuremath{\mat{B}}}
\providecommand{\mC}{\ensuremath{\mat{C}}}
\providecommand{\mD}{\ensuremath{\mat{D}}}
\providecommand{\mE}{\ensuremath{\mat{E}}}
\providecommand{\mF}{\ensuremath{\mat{F}}}

\providecommand{\mH}{\ensuremath{\mat{H}}}
\providecommand{\mI}{\ensuremath{\mat{I}}}

\providecommand{\mK}{\ensuremath{\mat{K}}}
\providecommand{\mL}{\ensuremath{\mat{L}}}

\providecommand{\mP}{\ensuremath{\mat{P}}}
\providecommand{\mQ}{\ensuremath{\mat{Q}}}

\providecommand{\mS}{\ensuremath{\mat{S}}}

\providecommand{\mU}{\ensuremath{\mat{U}}}
\providecommand{\mV}{\ensuremath{\mat{V}}}
\providecommand{\mW}{\ensuremath{\mat{W}}}
\providecommand{\mX}{\ensuremath{\mat{X}}}

\providecommand{\vb}{\ensuremath{\vec{b}}}

\providecommand{\ve}{\ensuremath{\vec{e}}}
\providecommand{\vf}{\ensuremath{\vec{f}}}

\providecommand{\vp}{\ensuremath{\vec{p}}}
\providecommand{\vq}{\ensuremath{\vec{q}}}

\providecommand{\vu}{\ensuremath{\vec{u}}}
\providecommand{\vw}{\ensuremath{\vec{w}}}
\providecommand{\vx}{\ensuremath{\vec{x}}}
\providecommand{\vy}{\ensuremath{\vec{y}}}
\providecommand{\vz}{\ensuremath{\vec{z}}}

\def\blue{\textcolor{black}}

\newcommand{\st}{{\rm s.t.}}

\newcommand{\m}{\boldsymbol}
\allowdisplaybreaks[4]
\usepackage[colorlinks = true,
linkcolor = blue,
urlcolor  = blue,
citecolor = blue,
anchorcolor = blue]{hyperref}
\renewcommand*{\thefootnote}{\fnsymbol{footnote}}
\newcommand{\mc}[1]{\mathcal{#1}}
\newcommand{\mbb}[1]{\mathbb{#1}}

\usepackage{amsthm}

\usepackage[usestackEOL]{stackengine}
\stackMath
\usepackage[framemethod=TikZ]{mdframed}
\usepackage{bm}
\mdfdefinestyle{MyFrame}{%
	linecolor=black,
	outerlinewidth=1.5pt,
	roundcorner=1.5pt,
	innerrightmargin=5pt,
	innerleftmargin=5pt,}

\usepackage{tikz}
\usetikzlibrary{matrix,positioning,decorations.pathreplacing}
\usetikzlibrary{shapes.geometric, arrows}
\usetikzlibrary{backgrounds}
\usetikzlibrary{shapes}
\usetikzlibrary{tikzmark}
\usetikzlibrary{calc}
\usetikzlibrary{arrows,shapes,positioning,shadows,trees,mindmap}
\usepackage[edges]{forest}
\usetikzlibrary{arrows.meta}
\colorlet{linecol}{black!75}
\usepackage{xkcdcolors} % xkcd colors

\tikzstyle{startstop} = [rectangle, rounded corners, minimum width=3cm, minimum height=1cm,text centered, draw=black, fill=red!10]
\tikzstyle{io} = [trapezium, trapezium left angle=70, trapezium right angle=110, minimum width=3cm, minimum height=1cm, text centered, text width=5cm, draw=black, fill=blue!10]
\tikzstyle{process} = [rectangle, minimum width=3cm, minimum height=1cm, text centered, draw=black, fill=orange!10]
\tikzstyle{decision} = [diamond, minimum width=2cm, minimum height=2cm, text centered, text width=3cm, draw=black, fill=green!10]
\tikzstyle{arrow} = [thick,->,>=stealth]

\newcolumntype{L}{>{\arraybackslash}m{3in}}
\newcolumntype{R}{>{\arraybackslash}m{1in}}
\newcolumntype{W}{>{\centering\arraybackslash}m{4in}}
\newcolumntype{K}{>{\centering\arraybackslash}m{2.2in}}

\usepackage{ltablex}
\usepackage[export]{adjustbox}
\usepackage{threeparttable}
\usepackage{multirow}
\definecolor{indiagreen}{rgb}{0.07, 0.53, 0.03}

\begin{document}
	\title{Comprehensive Framework for Controlling Nonlinear Multi-species Water Quality Dynamics}
	\author{Salma M. Elsheri$\text{f}^{\dagger, \P}$, Ahmad F. Tah$\text{a}^{\dagger, \ast \ast}$
		, Ahmed A. Abokif$\text{a}^{\diamond}$, and Lina Sel$\text{a}^{\spadesuit}$
		\vspace{-0.7cm}
		\thanks{$^\dagger$Department of Civil and Environmental Engineering, Vanderbilt University, Nashville, TN, USA. Emails: salma.m.elsherif@vanderbilt.edu, ahmad.taha@vanderbilt.edu}
		\thanks{$\P$Secondary: Department of Irrigation and Hydraulics Engineering, Faculty of Engineering, Cairo University.}
		\thanks{$^{\diamond}$Department of Civil, Materials, and Environmental Engineering, The University of Illinois Chicago. Email: abokifa@uic.edu}
		\thanks{$^{\spadesuit}$Department of Civil, Architecture, and Environmental Engineering, The University of Texas at Austin. Email: linasela@utexas.edu}
		\thanks{$^{\ast \ast}$Corresponding author.}
		\thanks{This work is partially supported by National Science Foundation under grants 1728629, 2015603, 2015671, 2151392, and 2015658.}
		}
	 
	\maketitle

%\markboth{...}{} 
\begin{abstract}
	Tracing disinfectant (e.g., chlorine) and contaminants evolution in water networks requires the solution of 1-D advection-reaction (AR) partial differential equations (PDEs). With the absence of analytical solutions in many scenarios, numerical solutions require high-resolution time- and space-discretizations resulting in large model dimensions. This adds complexity to the water quality control problem. In addition, considering multi-species water quality dynamics rather than the single-species dynamics produces a more accurate description of the reaction dynamics under abnormal hazardous conditions (e.g., contamination events). Yet, these dynamics introduce nonlinear reaction formulation to the model. To that end, solving nonlinear 1-D AR PDEs in real time is critical in achieving monitoring and control goals for various scaled networks with a high computational burden. In this work, we propose a novel comprehensive framework to overcome the large-dimensionality issue by introducing different approaches for applying model order reduction (MOR) algorithms to the nonlinear system followed by applying a real-time water quality regulation algorithm that is based on an advanced model to maintain desirable disinfectant levels in water networks under multi-species dynamics.  
	% These MOR approaches focus on reducing model dimension by orders of magnitude while preserving important properties of the original system (e.g., controllability). 
	The performance of this framework is validated using rigorous numerical case studies under a wide range of scenarios demonstrating the challenges associated with regulating water quality under such conditions.
\end{abstract}
	

\section{Introduction and Literature Review}~\label{sec:Into-Lit}
Water quality dynamics are widely modeled by the one-dimension advection-reaction (1-D AR) partial differential equations (PDEs). These AR PDEs allow the tracing of the disinfectant and other chemical substances' evolution throughout the components of water distribution networks (WDNs). In most cases, analytical solutions are non-existent to solve these PDEs network-wide. Nonetheless, PDEs can be solved using numerical techniques, yet, they require high-resolution time- and space-discretization. This results in high-dimension models that add computational burden to the problem of regulating water quality in drinking networks. That leads to physical-driven models that are intractable when considering constrained control and water quality (WQ) regulation algorithms. 

Moreover, in water quality simulations the most widely used decay and reaction model is the single-species model. In this model, disinfectant (i.e., chlorine) is assumed to decay at a constant rate that only accounts for purified water contamination levels. Yet, contamination sources vary from microbial, non-microbial components in the bulk flow, attached to the pipe walls, or contamination events that get intruded into the system \cite{palansooriya2020occurrence}. This drives the need for a more accurate representation of these scenarios which can be achieved by the multi-species reaction dynamics. The multi-species dynamics enable the model to simulate chlorine evolution with the existence of another reactive component in the system. This representation duplicates the number of variables to be traced network-component-wide while unfortunately adding complexity to the model by introducing nonlinear reaction dynamics.

To this end, model order reduction (MOR) is an essential step to move forward in achieving a compact formulation of the multi-species water quality dynamics to be integrated into a model-based control framework. MOR techniques transform the \textit{full-order model (FOM)} to a \textit{reduced-order model (ROM)} in a way that preserves the structure, properties, and the closed-form representation of the FOM while achieving the pre-specified level of accuracy and reducing computational time. Eventually, the goal is to control chlorine injections dosed by rechlorination stations to maintain residual levels that meet water quality standards. That can be achieved by applying an effective control algorithm on the derived ROM.

Our group has been interested in various dimensions of this research area. A summary of our work and the prior literature is given next. 

\subsection{Literature Review}~\label{sec:LitRev}
Hereinafter, we survey the literature on the topics of MOR for dynamic systems, in general, and water systems, in particular, and water quality regulation and control while highlighting the gaps and drawbacks motivating this paper's contributions.  

\textit{MOR for Dynamic Systems:} Several studies have proposed and implemented different MOR algorithms in various disciplines (e.g., electromagnetics, electro-mechanics, structural and fluid dynamics) where the large-dimensionality issue is faced \cite{moorePrincipalComponentAnalysis1981,baurModelOrderReduction2014,montier2017balanced,rutzmoser2018model}. Most of these studies have applied either singular value decomposition (SVD) approach \cite{rowleyMODELREDUCTIONFLUIDS2005,willcox2002balanced} or Krylov subspace methods \cite{grimme1997krylov,beattie2008interpolation}. Whilst, combining SVD and Krylov methods is investigated and implemented in \cite{gugercin2008iterative}. In infrastructure networks preserving the system's properties including stability, controllability, and observability is a major concern while applying MOR with the aim of applying a post-reduction control. Nevertheless, Krylov methods do not preserve such properties which limit their suitability to be used in our study \cite{baurModelOrderReduction2014}.

Several SVD-based model reduction methods have been proposed for linear systems and more realizations and extensions have been investigated and integrated to tackle the reduction of nonlinear systems; a review of linear and nonlinear models order reduction can be found in \cite{antoulas2000survey,kumar2022state}. The Balanced Truncation (BT) method \cite{moorePrincipalComponentAnalysis1981} is built based on both the controllability and observability Gramians for stable, linear systems. The study by \cite{lall2002subspace} extends BT to be applied to nonlinear systems, while the authors in \cite{barrachina2005parallel,zhou1999balanced} build extensions for unstable systems. However, BT becomes computationally intractable for large-scale systems. Nevertheless, the famous widely used Proper Orthogonal Decomposition (POD) in fluid dynamics community \cite{sirovich1987turbulence} is considered tractable at the expense of accuracy compared to BT. Yet, in some cases where relatively lower accuracy is acceptable, POD may result in an unstable system even near stable equilibrium points depending on the actual formulation of the full-order model. Therefore, the methods that balance between the BT and POD methods have been propounded to integrate the advance of both methods into one. For example, the authors in \cite{willcoxBalancedModelReduction2002} have proposed a balanced method but it has failed to successfully reduce models when the number of outputs of the system is large. Conversely, the balanced POD (BPOD) \cite{rowleyMODELREDUCTIONFLUIDS2005} is tractable with an overall computation time similar to POD, but it computes adjoint snapshots to combine and balance controllability and observability similarly to BT, which is not raised in POD. Furthermore, POD can be extended to reduce the order of nonlinear systems by approximately projecting the nonlinearity term in the system to a subspace of the dynamics \cite{nguyen2020pod,baurModelOrderReduction2014}. Therefore, the nonlinear term is evaluated separately and approximated 
%However, the evaluation for the linear system can pre-computed only once at the offline stage but the nonlinear one needs to be reassessed more frequently to capture the nonlinearity behavior and work within lower dimension space. Nonetheless, high frequency at each time step is impractical as it depends on the large dimension of the full original model for actual representation. This makes the computation time for the nonlinear reduced model as expensive as that of the full original system. 
at only a small set of interpolation points (hyper-parameter) using a combination of projection and interpolation methods such as the discrete empirical interpolation method (DEIM) \cite{nguyen2020pod}, the Gappy POD method \cite{galbally2010non,akkari2019novel}, and the Gauss-Newton with approximated tensors (GNAT) method \cite{carlberg2013gnat}, refer to the review paper \cite{benner2015survey} for details.

\textit{MOR for Water Systems:} Water systems model order reduction has been broadly investigated for network hydraulics over the past decades with a limited number of studies looking into MOR for water quality dynamics. These studies adopt different approaches to reduce the hydraulic model dimension by applying methods varying between performing nodal Gaussian elimination \cite{ulanicki1996simplification}, Gaussian elimination on the linearized form of the model and recovering the nonlinearity of the system as a post-reduction step \cite{martinez2014fast}, genetic algorithm \cite{shamirOptimalRealTimeOperation2008}, and system aggregation \cite{preis2011efficient}. Perelman and Ostfeld \cite{perelmanWaterDistributionSystem2008} consider a coupled model that combines both hydraulics and water quality dynamics of the network and apply systems aggregation.

Lately, two studies have applied different approaches to cover the MOR for water quality dynamics. Authors in \cite{elkhashapModelOrderReduction2022} have proposed reducing the order of the water quality model by formulating a bi-linear spatially-discretized but a temporally-continuous representation of the dynamics. This formulation augments the input vectors in a way that preserves the system's stability. The induced error between the actual and reduced-order models is minimized by the reduction of the $\mathcal{H}_2$-norm. In this study, water dynamics transport and reaction are simulated using the Advection-Diffusion-Reaction Partial Differential Equations that include the diffusion term in comparison to our work that neglects its effect. Nonetheless, studies \cite{li2005importance,shang2021lagrangian} state that diffusion is dominated in network branches with significantly low velocities. To that end, it is an acceptable assumption to neglect the diffusion effect in networks with limited dead-end branches, higher velocities, and changing demands. On the other hand, augmenting and transferring the model into nonlinear formulation results in a more complex one when considering the multi-species nonlinear water quality dynamics and does not preserve the stability of the system.

\begin{figure*}[t!]
	\includegraphics[width=\textwidth]{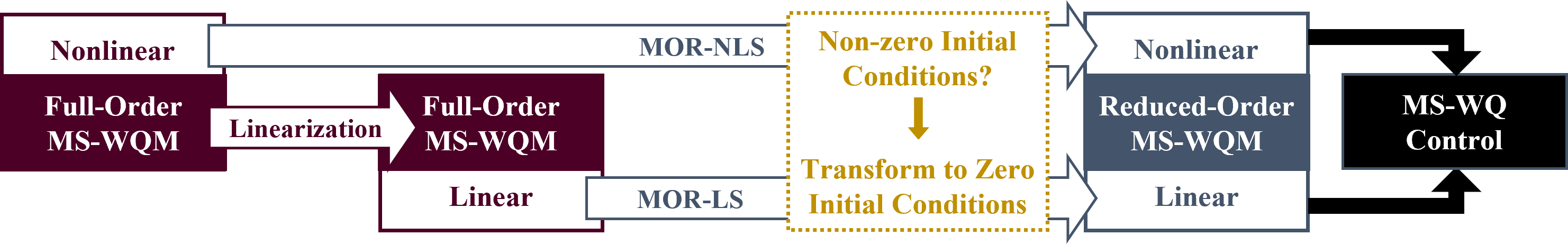}
	\caption{Conceptual framework of the paper.}~\label{fig:FW} \vspace{-0.25cm}
\end{figure*}

Secondly, study \cite{wang2022model} applies different SVD-based projection algorithms to reach a reduced-order water quality model including BT, POD, and BPOD in addition to preserving the stability of the BPOD method. Results have proven that the BPOD method is more usable while being computationally tractable and robust for zero and non-zero initial conditions. However, their model only includes single-species linear reaction dynamics where chlorine is assumed to be decaying at a constant rate resulting in a linear state-space formulation. Therefore, our work allows filling the gap in applying MOR for multi-species nonlinear dynamics. Moreover, in their model the explicit central Lax-Wendroff discretization scheme is used. However, the Upwind schemes give a more accurate physical description of the advection-reaction problem. In our study, we apply the \textit{explicit} and \textit{implicit} Upwind discretization schemes while highlighting the differences and the level of difficulty. Notice that, on the contrary to studies \cite{fuModelingWaterQuality2020,lassilaModelOrderReduction2014a,he2014reduced} where MOR is performed for compositional simulation, the authors in \cite{wang2022model} state that it is considered as a pre-step to apply an efficient control algorithm which also applies to our work in this paper. 

\textit{Water Quality Control:} The topic of controlling chlorine has been covered in several studies with various algorithms, objectives, and constraints \cite{ohar2014optimal,ostfeld2006conjunctive,munavalli2003optimal}. Objectives vary between minimizing the cost of injecting chlorine into the system, maintaining minimal deviations from chlorine setpoint concentrations, minimizing the formation of the excess DBPs, and minimizing computational time  \cite{fisher2018framework}. The problem formulation is either a single-objective optimization problem or a multi-objective one with more of the aforementioned objectives. However, such studies do not build a closed-form representation of all inputs, states, and outputs that updates every specified time-step over the simulation period and allows network-wide control. Whereas, studies \cite{wang2021effective,wang2022model} apply Model Predictive Control (MPC) on the full-order and reduced-order single-species models in both studies with no clear explanation/extension for scenarios where multi-species dynamics take place. 

\textit{Our prior Work:} We have been focusing on tackling and covering the water quality modeling and control in WDNs. First, the problem of modeling and controlling single-species water quality dynamics is thoroughly investigated in \cite{wang2021effective}, followed by reducing this model's order and verifying the validity of controlling the reduced order model in \cite{wang2022model}. Moreover, as a first state-of-the-art attempt, study \cite{wangSystemId2023} has identified single-species water quality models using only input-output experimental data and, accordingly, data-driven system identification algorithms. Lastly, a survey study on how to accurately simulate multi-species water quality dynamics has been conducted in \cite{ELSHERIF2022}. This study has built a closed-form, network- and control-theoretic representation of all system inputs, variables, and output measurements under such dynamics that give a more realistic WQ formulation. The performance of this formulation has been validated using the widely-used simulation tool, EPANET and its multi-species water quality simulation extension, EPANET-MSX \cite{rossman2020epanet,shangEPANETMSXUserManual2023}. However, controlling chlorine under multi-species dynamics, based on a control-theoretic explicit model, is to the authors' knowledge has not been investigated---a gap that is filled in this paper.

%One more drawback of their study, investigating the effect of changing the hydraulics states for the same water network is not covered. 

\subsection{Paper Contributions}~\label{sec:PaperContrb}
This paper's major objective is to investigate the implementation and complexity of regulating and controlling chlorine levels under multi-species water quality. The detailed paper contributions are: 
\begin{itemize}
	\item Construct and propose a comprehensive framework to overcome the large-dimensionality issue associated with discretizing the 1-D AR PDEs and the complexity associated with the nonlinearity of the multi-species water quality dynamics. Different paths can be taken, starting by linearizing the system and applying MOR for linear systems (MOR-LS). Another path is to consider the nonlinear MOR (MOR-NLS) algorithm on the original FOM. 
	\item Utilizing the reduced-order models in an MPC algorithm. We apply it to the formulated ROMs and compare them to each other and to the original FOMs. Also, we compare it with basic scenarios with single-species dynamics demonstrating the challenges associated with controlling chlorine levels under multi-species water quality dynamics. 
	\item Position the framework in a generalized scalable form in the sense that simplifications are included to consider single-species water dynamics and differentiations are suggested to consider chlorine linear/nonlinear decay and reaction models that have been developed in the literature to simulate various events/scenarios. 
	\item Validate the performance of the framework using thorough numerical case studies to test accuracy, computational burden, and robustness to the system hydraulics changes.
\end{itemize}

Our proposed framework is illustrated in Fig. \ref{fig:FW}. As shown, different approaches can be followed to formulate a reduced-order model to be controlled for the multi-species water quality model. Each step to be taken and each path to be chosen are explained in the following sections of the paper. The paper's sections are organized as follows, Section "\hyperref[sec:WQModel]{State-space Multi-species Water Quality Model}" provides the formulation of the state-space representation of the multi-species water quality model (MS-WQM). This formulation is based on the transport and reaction model in pipes, mass balance for the other network components, and the multi-species dynamics expression. Section "\hyperref[sec:Methods]{Model Order Reduction and Transformation of MS-WQM}" provides full descriptions of the methods used in our framework to reach a compact reduced-order model. Section "\hyperref[sec:MS-MPC]{Real-Time Regulation of MS-WQM via Model Predictive Control and McCormick Relaxations}" introduces the control problem and its implementation on the linear and nonlinear ROM. Section "\hyperref[sec:CaseStud]{Case Studies}" showcases the framework performance on different networks under a wide range of scenarios. Section "\hyperref[ConcLimRoc]{Conclusion, Paper's Limitations, and Recommendations for Future Work}" comes last.

\section{State-space Multi-species Water Quality Model}~\label{sec:WQModel}
We model WDN by a directed graph $\mathcal{G} = (\mathcal{N},\mathcal{L})$.  The set $\mathcal{N}$ defines the nodes and is partitioned as $\mathcal{N} = \mathcal{J} \cup \mathcal{T} \cup \mathcal{R}$ where sets $\mathcal{J}$, $\mathcal{T}$, and $\mathcal{R}$ are collections of junctions, tanks, and reservoirs. Let $\mathcal{L} \subseteq \mathcal{N} \times \mathcal{N}$ be the set of links, and define the partition $\mathcal{L} = \mathcal{P} \cup \mathcal{M} \cup \mathcal{V}$, where sets $\mathcal{P}$, $\mathcal{M}$, and $\mathcal{V}$ represent the collection of pipes, pumps, and valves. Total number of states is $n_x=n_L+n_N$, where $n_\mathrm{L}$ and $n_\mathrm{N}$ are numbers of links and nodes. The number of reservoirs, junctions, tanks, pumps, valves, and pipes are $n_\mathrm{R}, n_\mathrm{J}, n_\mathrm{TK}, n_\mathrm{M}, n_\mathrm{V},$ and $n_\mathrm{P}$. Each pipe $i$ with length $L_i$ is spatially discretized and split into $s_{L_i}$ segments. Hence, number of links is expressed as $n_\mathrm{L}=n_\mathrm{M}+n_\mathrm{V}+ \sum_{i=1}^{n_\mathrm{P}} s_{L_i}$ while $n_\mathrm{N}=n_\mathrm{R}+n_\mathrm{J}+n_\mathrm{TK}$ is the number of nodes.

In this paper, the state-space representation is formulated for multi-species dynamics with two chemicals: chlorine and a fictitious reactant. The system representation of the two-species which is able to capture chemicals evolution, booster stations injections, and sensors measurements, is expressed by an NDE as follows

\begin{subequations}~\label{equ:NDE}
	\begin{align}
		\begin{split}
			\underbrace{\begingroup % keep the change local
				\setlength\arraycolsep{2pt}	\begin{bmatrix}
					\mE_{11}(t) & 0 \\ 0 & \mE_{22}(t)
				\end{bmatrix} \endgroup}_{{\mE}(t)} 
			\underbrace{	\begin{bmatrix}
					\vx_1(t+\Delta t) \\ \vx_2(t+\Delta t)
			\end{bmatrix}}_{{\vx}(t+\Delta t)} & = 
			\underbrace{\begingroup % keep the change local
				\setlength\arraycolsep{2pt}	\begin{bmatrix}
					\mA_{11}(t) & 0 \\ 0 & \mA_{22}(t)
				\end{bmatrix} \endgroup}_{{\mA}(t)}
			\underbrace{ \begin{bmatrix}
					\vx_1(t) \\ \vx_2(t)
			\end{bmatrix}}_{{\vx}(t)} +  \underbrace{\begingroup % keep the change local
				\setlength\arraycolsep{2pt}	\begin{bmatrix}
					\mB_{11}(t) & 0 \\ 0 & \mB_{22}(t)
				\end{bmatrix} \endgroup}_{{\mB}(t)} 
			\underbrace{\begin{bmatrix}
					\vu_1(t) \\ \vu_2(t)
			\end{bmatrix}}_{{\vu}(t)} + \vf(\vx_1,\vx_2,t), \label{equ:NDE1}
		\end{split}
		\\ \begin{split}
			\underbrace{	\begin{bmatrix}
					\vy_1(t) \\ \vy_2(t)
			\end{bmatrix}}_{{\vy}(t)} = 
			\underbrace{	\begin{bmatrix}	\mC_{11}(t) & 0 \\
					0 & \mC_{22}(t)
			\end{bmatrix}}_{{\mC}(t)} &
			\underbrace{ \begin{bmatrix}
					\vx_1(t) \\ \vx_2(t)
			\end{bmatrix}}_{\vx(t)} + \underbrace{	\begin{bmatrix}	\mD_{11}(t) & 0 \\
					0 &  \mD_{22}(t)
			\end{bmatrix}}_{{\mD}(t)}
			\underbrace{ \begin{bmatrix}
					\vu_1(t) \\ \vu_2(t)
			\end{bmatrix}}_{\vu(t)}\label{equ:NDE2}
		\end{split}
	\end{align}
\end{subequations}
where variable $t$ represents specific time in a simulation period $[0,T_s]$; $\Delta t$ is the time-step or sampling time; vectors $\vx_1(t)$ and $\vx_2(t) \in \mathbb{R}^{n_x}$ depict  the concentrations of chlorine and the other fictitious reactant (two species model) in the entire network; vector $\vu_1(t) \in \mathbb{R}^{n_{u_1}}$ represents the dosages of injected chlorine; vector $\vu_2(t) \in \mathbb{R}^{n_{u_2}}$ accounts for planned or unplanned injection of the fictitious component; vector $\vf(\vx_1,\vx_2,t)$ encapsulates the nonlinear part of the equations representing the mutual nonlinear reaction between the two chemicals; vector $\vy_1(t) \in \mathbb{R}^{n_{y_1}}$ denotes the sensor measurements of chlorine concentrations at specific locations in the network while $\vy_2(t) \in \mathbb{R}^{n_{y_2}}$ captures the fictitious reactant measurements by sensors in the network if they exist. The state-space matrices $\{\mE, \mA, \mB, \mC, \mD\}_{\bullet}$ are all time-varying matrices that depend on the network topology and parameters, hydraulic parameters, decay rate coefficients for the disinfectant, and booster stations and sensors locations. It is customary to assume that these matrices evolve at a slower pace than the states $\vx(t)$ and control inputs $\vu(t)$. On another note, matrices $\mE_{11}, {\mE}_{22}$ are changing every hydraulic time-step allowing them to be represented at time $t$ not $t + \Delta t$ of the water quality simulation horizon.

The concentration evolution throughout network components is covered by the conservation of mass law, transport, decay, and reaction models of the substances. A full description of how the models are derived for each type of the components is provided in \cite{ELSHERIF2022}. However, for the reader to be able to follow up with the developments of this paper, some material from \cite{ELSHERIF2022} need to be reproduced and altered. We list a brief overview of the governing equations formulating our model and its state-space representation in the following sections.

\subsection{Transport and Reaction in Pipes}~\label{sec:TranReactPipes}
Conservation of mass during transport and reaction in pipes is simulated by the one-dimension advection-reaction (1-D AR) partial differential equation, which for Pipe $i$ is expressed as
\begin{equation}\label{equ:PDE}
	\partial_t c_i^\mathrm{P}=-v_i(t) \partial_t c_i^\mathrm{P} + R^\mathrm{P}_{\mathrm{MS}}(c_i^\mathrm{P}(x,t)),
\end{equation}
where $c^\mathrm{P}_i(x,t)$ is concentration in pipe at location $x$ and time $t$; $v_i(t)$ is the mean flow velocity; and $R^\mathrm{P}_{\mathrm{MS}}(c^\mathrm{P}_i(x,t))$ is the multi-species reaction rate in pipes expression (more explanation is given in Section "\nameref{sec:MSmodel}").

\begin{figure}[h!]
	\centering
	\includegraphics[width=0.7\textwidth]{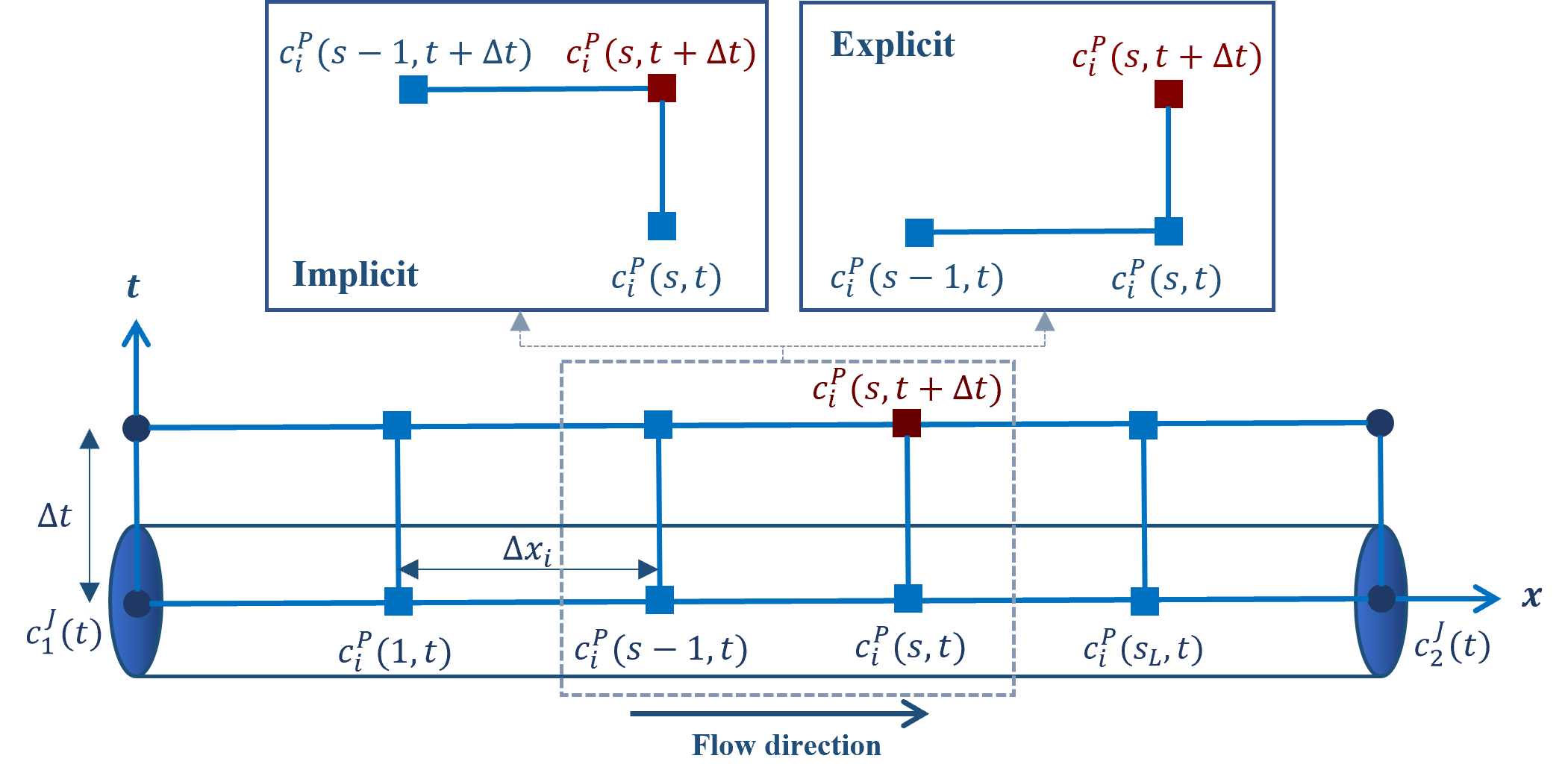}
	\caption{Implicit and Explicit Upwind discretization schemes for Pipe $i$ connecting Junctions 1 and 2. Each scheme calculates concentration $c^\mathrm{P}_i(s,t+\Delta t)$ at segment $s$ (colored in maroon) depending on concentrations at the segments/nodes included in its frame. (For interpretation of the references to color in this figure legend, the reader is referred to the web version of this article.)}~\label{fig:UW}
\end{figure}

Eq. \eqref{equ:PDE} is discretized over a fixed spatio-temporal grid, that for a Pipe $i$ with length $L_{i}$ is split into a number of segments $s_i=\Big\lfloor \frac{L_i}{v_i(t) \Delta t} \Big\rfloor$ of length $\Delta x_i= \frac{L_i}{s_i}$. In the considered 1-D AR model, the main two processes are the advection where the concentration at a certain location and time is affected by upstream concentrations, and reaction where chemicals decay and/or mutually react. That being said, \textit{Upwind} discretization schemes are more descriptive to the actual physical process considered among other schemes \cite{hirsch1990numerical}. Applying the Eulerian Finite-Difference based Implicit Upwind scheme on the multi-species water quality dynamics representation adapted in this paper has shown reliable results that trace chemicals contractions within different networks with various scales, according to \cite{ELSHERIF2022}. In this paper we consider both \textit{Explicit} and \textit{Implicit} Upwind schemes to investigate their performance from a control-theoretic perspective (See Fig. \ref{fig:UW}). 

\subsubsection{Explicit Upwind Scheme} 
For segment $s$ of Pipe $i$ except for the first segment, the concentration is calculated as
\begin{equation}\label{equ:EUW_Si}
	c^\mathrm{P}_i(s,t+ \Delta t) = (1-{\lambda}_i(t)) c^\mathrm{P}_i(s,t) +{\lambda}_i(t) c^\mathrm{P}_i(s-1,t)+R^\mathrm{P}_{\mathrm{MS}}(c^\mathrm{P}_i(s,t)) \Delta t,
\end{equation}
where ${\lambda}_i(t)=\frac{v_i(t) \Delta t}{\Delta x_i}$ is the Courant number and according to Courant-Friedrichs-Lewy condition (CFL), Courant number (CN) is maintained to be in the range of  $0<{\lambda}_i(t) \leq 1$ so that the scheme is stable. Moreover, the concentrations in the first segment is expressed as in \eqref{equ:EUW_S1} assuming that the connecting upstream node is Junction $j$.
\begin{equation}\label{equ:EUW_S1}
	c^\mathrm{P}_i(1,t+ \Delta t) = (1-{\lambda}_i(t)) c^\mathrm{P}_i(1,t) +{\lambda}_i(t) c^\mathrm{J}_j(t)+R^\mathrm{P}_{\mathrm{MS}}(c^\mathrm{P}_i(s,t)) \Delta t,
\end{equation}

\subsubsection{Implicit Upwind Scheme}
The difference is that the concentration at the upstream segment/node is taken at the current time-step we are calculating at. That is, Equ. \eqref{eq:IUW_Si} calculates the concentration for segment $s$ of Pipe $i$. As well, the concentration of the first segment with Junction $j$ as the upstream node is expressed in Equ. \eqref{eq:IUW_S1}.

\begin{subequations}
	\begin{align}
		&	(1+{\lambda}_i(t))c^\mathrm{P}_i(s,t+ \Delta t) - {\lambda}_i(t) c^\mathrm{P}_i(s-1,t+ \Delta t) = c^\mathrm{P}_i(s,t)+R^\mathrm{P}_{\mathrm{MS}}(c^\mathrm{P}_i(s,t)) \Delta t,~\label{eq:IUW_Si}\\
		&	(1+{\lambda}_i(t))c^\mathrm{P}_i(1,t+ \Delta t) - {\lambda}_i(t) c^\mathrm{J}_1(t+ \Delta t) = c^\mathrm{P}_i(1,t)+R^\mathrm{P}_{\mathrm{MS}}(c^\mathrm{P}_i(1,t)) \Delta t.~\label{eq:IUW_S1}
	\end{align}
\end{subequations}

\subsection{Mass Balance at Network Components}
For components other than pipes, conservation of mass is applied to formulate expressions for concentrations calculation. 

\subsubsection{Mass Balance at Reservoir} Reservoirs are assumed to have constant concentrations. For each Reservoir $i$ concentration is expressed as $c_i^\mathrm{R}(t+\Delta t)=c_i^\mathrm{R}(t).$

\subsubsection{Mass Balance at Pumps and Valves} The model deals with pumps and valves as transmission links with concentration equals the concentration of the node upstream them. That being said, for Pump $i$ or Valve $j$ installed after Reservoir $k$ (as an example), concentrations are expressed as $c_i^{\mathrm{M}}(t+\Delta t) = c_k^{\mathrm{R}}(t+\Delta t),$ and $c_j^{\mathrm{V}}(t+\Delta t) = c_k^{\mathrm{R}}(t+\Delta t).$

\subsubsection{Mass Balance at Junctions} Chemicals are assumed to have complete and instantaneous mixing in junctions with no storage time. Thus, chemical concentration at each Junction $i$ is expressed as
\begin{equation}~\label{equ:mb-junc} 
	c_i^\mathrm{J}(t)= \frac{\sum_{j \in L_{\mathrm{in}}} q_{\mathrm{in}}^{j}(t) c_\mathrm{in}^j(t)+q^\mathrm{B_\mathrm{J}}_i(t) c^\mathrm{B_\mathrm{J}}_i(t)}{q^{\mathrm{D}_\mathrm{J}}_i(t)+\sum_{k \in L_{\mathrm{out}}} q_{\mathrm{out}}^{k}(t)},
\end{equation}
where $j$ and $k$ are the counters for total $L_{\mathrm{in}}$ links flowing into the junction and $L_{\mathrm{out}}$ links extracting flow from the junction; $q_{\mathrm{in}}^{j}(t)$ and $q_{\mathrm{out}}^{k}(t)$ are the inflows and outflows from these links connected to the junction; $c_\mathrm{in}^j(t)$ is the concentration in the inflow solute;  $q^\mathrm{B_\mathrm{J}}_i(t)$ is the flow injected to the junction with concentration $c^\mathrm{B_\mathrm{J}}_i(t)$ by booster station if located; and $q^{\mathrm{D}_\mathrm{J}}_i(t)$ is demand.

\subsubsection{Mass Balance at Tanks} Mass conservation in tanks assumes complete instantaneous mixing of all inflows, outflows, and stored water following the continuously stirred tank reactor (CSTR) model. 
\begin{equation}\label{equ:tank2}
	\begin{split}
		V_i^\mathrm{TK}(t + \Delta t) c_i^\mathrm{TK}(t+ \Delta t) = V_i^\mathrm{TK}(t) c_i^\mathrm{TK}(t) 
		& +\sum_{j \in L_{\mathrm{in}}} q^j_\mathrm{in}(t)c^j_\mathrm{in}(t) \Delta t
		+V^\mathrm{B_\mathrm{TK}}_i(t+\Delta t)c^\mathrm{B_\mathrm{TK}}_i(t+\Delta t)\\
		& - \sum_{k \in L_{\mathrm{out}}} q^k_\mathrm{out}(t)c_i^\mathrm{TK}(t) \Delta t
		+R^\mathrm{TK}_{\mathrm{MS}}(c_i^\mathrm{TK}(t)) V_i^\mathrm{TK}(t) \Delta t,
	\end{split}
\end{equation}
where $V^\mathrm{B_\mathrm{TK}}_i(t+\Delta t)$ is the volume injected to the tank with concentration $c^\mathrm{B_\mathrm{TK}}_i(t+\Delta t)$ by booster station if located. $R^\mathrm{TK}_{\mathrm{MS}}(c^\mathrm{P}_i(x,t))$ is the multi-species reaction rate in tanks expression (refer to Section "\nameref{sec:MSmodel}").

\subsection{Multi-species Reaction and Decay Model}~\label{sec:MSmodel} Dividing the model into decay and mutual reaction dynamics allows it to consider a substance with relatively different reaction rates than the decay rate and for the model to be less sensitive to the other reactants' concentrations. Decay model is a first-order model that depends on only chlorine concentration and constant decay rate. Hence, the chlorine decay reaction rates for Pipe $i$ and Tank $j$ are $k_i^\mathrm{P} = k_{b}+\frac{2k_{w}k_{f}}{r_{\mathrm{P}_i}(k_{w}+k_{f})},\,\,\,\,\, k_j^\mathrm{TK} = k_{b}$, where $ k_{b}$ is the bulk reaction rate constant; $k_{w}$ is the wall reaction rate constant; $k_{f}$ is the mass transfer coefficient between the bulk flow and the pipe wall; $r_{\mathrm{P}_i}$ is the pipe radius.\\
The mutual reaction model is expressed by a second-order nonlinear ODEs which are discretized using Forward Euler method  
$c(t+\Delta t)-c(t) =-k_r \Delta t(c(t) \tilde{c}(t)), \;
\tilde{c}(t+\Delta t)-\tilde{c}(t) =-k_r \Delta t(c(t)\tilde{c}(t)),$ where  $c(t), \tilde{c}(t)$ are the concentrations for chlorine and fictitious reactant; and $k_r$ is the mutual reaction rate between them. Eventually, reaction expressions for pipes and tanks are 
\begin{subequations} ~\label{equ:MR_PTK}
	\begin{align}
		R^{\mathrm{P}}_{\mathrm{M}}(c^\mathrm{P}_i(s,t))=-k_r c^\mathrm{P}_i(s,t) \tilde{c}^\mathrm{P}_i(s,t), \;\;\; R^{\mathrm{TK}}_{\mathrm{M}}(c^\mathrm{TK}_j(t))=-k_r \tilde{c}^\mathrm{TK}_j(t) c^\mathrm{TK}_j(t),\\
		R^{\mathrm{P}}_{\mathrm{M}}(\tilde{c}^\mathrm{P}_i(s,t))=-k_r c^\mathrm{P}_i(s,t) \tilde{c}^\mathrm{P}_i(s,t), \;\;\; R^{\mathrm{TK}}_{\mathrm{M}}(\tilde{c}^\mathrm{TK}_j(t))=-k_r \tilde{c}^\mathrm{TK}_j(t) c^\mathrm{TK}_j(t).
	\end{align}
\end{subequations}

A full description of the state-space matrices construction for the Upwind discretization schemes and an example on a simple three-node network (consists of reservoir, a pump, a junction, a pipe, and a tank---Fig. 6) is included in \cite{ELSHERIF2022} for reader's reference on how to formulate the representation for different network component. {It is worth mentioning that this study  validates the utilization of these EFD discretization schemes and the model performance as mentioned in comparison to EPANET and its extension, EPANET-MSX (WQ multi-species simulation tool). The comparison is considered reliable as the governing laws and equations are the same for all network components in both models. It should be noted that EPANET+EPANET-MSX employs the Lagrangian time-driven method, dividing each pipe into changing-sized segments. Whilst, the adopted EFD schemes in our study work within a fixed grid, facilitating the construction of a state-space representation with finite dimensions. The drawback associated with these discretization schemes is the large dimensionality of the model. However, main objectives of this study are to address this challenge by employing model order reduction techniques and to integrate the reduced-order multi-species model effectively into time-efficient real-time feedback control algorithm, which are outlined and presented in detail in the next sections. On the other hand, coupling the EPANET+EPANET-MSX model with a real-time control algorithm is complex and presents challenges due to the need to handle changes in segment count and size per pipe at each simulation time-step, as well as being familiar with and able to leverage and use their toolkits in the coding language used (i.e., MATLAB and Python).

In the next section, we investigate different MOR algorithms for \eqref{equ:NDE}.  

\section{Model Order Reduction and Transformation of MS-WQM}~\label{sec:Methods}
The state-space representations formulated in the previous section are in forms of nonlinear difference equations (NDEs) \eqref{equ:NDE} with large numbers of variables resulted from high resolution spatio- temporal-discretization. To reach the end-goal of this paper, which is controlling chlorine levels for \eqref{equ:NDE}, we propose different methodologies to reduce the model order and showcase their limitations, accuracy, computational time, and robustness/sensitivity to initial conditions and fictitious reactant type. That being said, we list full descriptions of the methods covered in our framework. We start with linearizing \eqref{equ:NDE}, then explain model order reduction and transformation for linearized and original nonlinear systems. 

\subsection{Model Linearization}~\label{sec:SysLin}
%The state-space representation is formulated depending on transport, decay and reaction of the chemicals in the system. The mutual reaction dynamics in is in a form of nonlinear term adding more level of difficulty to the model. For pipes, the fourth technique (MoCs) studied in this paper is built on an assumption that allows for a simplification to formulate a linear system (explained in Section \ref{sec:MutualReact}). However, t That is, hereinafter we summarize the linearization process when considering one of the three techniques and also for tanks simulation.  
The mutual reaction is expressed as a nonlinear term that can be linearized using Taylor series approximations \cite{apostol1991calculus}. 
By linearizing around operating points $c_o, \tilde{c}_o$, the nonlinear term $R_\mathrm{M}(c(t),\tilde{c}(t))$ for both chemicals is expressed as:
\begin{equation} \label{eq:LDE}
	\begin{aligned}
		R_\mathrm{M}(c(t),\tilde{c}(t))&=-k_r (c_o \tilde{c}_o +c_{o}(\tilde{c}(t)-\tilde{c}_{o})+\tilde{c}_{o}(c(t)-c_{o})),\\
		&=-k_r (c_{o}\tilde{c}_{o}+c_{o}\tilde{c}(t)-c_{o}\tilde{c}_{o}+\tilde{c}_{o}c(t)-\tilde{c}_{o}c_{o}),\\
		&=-k_r (c_{o}\tilde{c}(t)+\tilde{c}_{o}c(t)-\tilde{c}_{o}c_{o}).
	\end{aligned}
\end{equation}
\indent For each of the chemicals, the mutual reaction after linearization is broken down to a term that depends on its concentration, a term that depends on the other chemical's concentration, and a constant. The general state-space representation \eqref{equ:NDE} has a block-diagonal matrix of $\mA$ matrices with no dependency between the chemical except in the $\vf$ function. That is, by applying linearization to the model the state-space representation is updated to linear difference equations (LDEs):
%\begin{equation} 
\begin{equation}\label{systemstatesLDE}
	\underbrace{
		%				\begin{bmatrix}
			\begingroup % keep the change local
			\setlength\arraycolsep{2pt}
			\begin{bmatrix}
				{\mE}_{11}(t) & 0 \\ 0 & {\mE}_{22}(t)
				%		\end{bmatrix}
		\end{bmatrix}
		\endgroup
	}_{{\mE}(t)} \underbrace{	\begin{bmatrix}
			\vx_1(t+\Delta t) \\ \vx_2(t+\Delta t)
	\end{bmatrix}}_{{\vx}(t+\Delta t)} = 
	\underbrace{ \begingroup % keep the change local
		\setlength\arraycolsep{2pt}
		\begin{bmatrix}
			{\breve{\mA}_{11}(t)} & {\breve{\mA}_{12}(t)} \\ {\breve{\mA}_{21}(t)} & {\breve{\mA}_{22}(t)}
		\end{bmatrix}
		\endgroup}_{{\breve{\mA}(t)}}
	\underbrace{ \begin{bmatrix}
			\vx_1(t) \\ \vx_2(t)
	\end{bmatrix}}_{{\vx}(t)}
	+ \underbrace{	\begin{bmatrix}
			\mB_{11}(t) & 0 \\ 0 & \mB_{22}(t)
	\end{bmatrix}}_{{\mB}(t)} 
	\underbrace{\begingroup % keep the change local
		\setlength\arraycolsep{1.5pt}
		\begin{bmatrix}
			\vu_1(t) \\ \vu_2(t)
		\end{bmatrix}
		\endgroup}_{{\vu}(t)} + {\m{\Phi}},
\end{equation}
%\end{equation}
where $\breve{\mA}_{11}(t)$ and $\breve{\mA}_{22}(t)$ are the modified diagonal matrices; $\breve{\mA}_{12}(t)$ and $\breve{\mA}_{21}(t)$ are the matrices gathering the dependency between the two species concentrations; and $\m\Phi$ is the vector containing the constants. Note that the changes in $\breve{\mA}_{11}(t)$ and $\breve{\mA}_{22}(t)$ from the original matrices are only in the sub-matrices/elements representing pipes and tanks only (i.e., ${\mA}^\mathrm{P}_\mathrm{P}$ and ${\mA}^\mathrm{TK}_\mathrm{TK}$).

\begin{figure}[t!]
	\centering	\subfloat[\label{fig:MOR1}]{\includegraphics[keepaspectratio=true,scale=0.5]{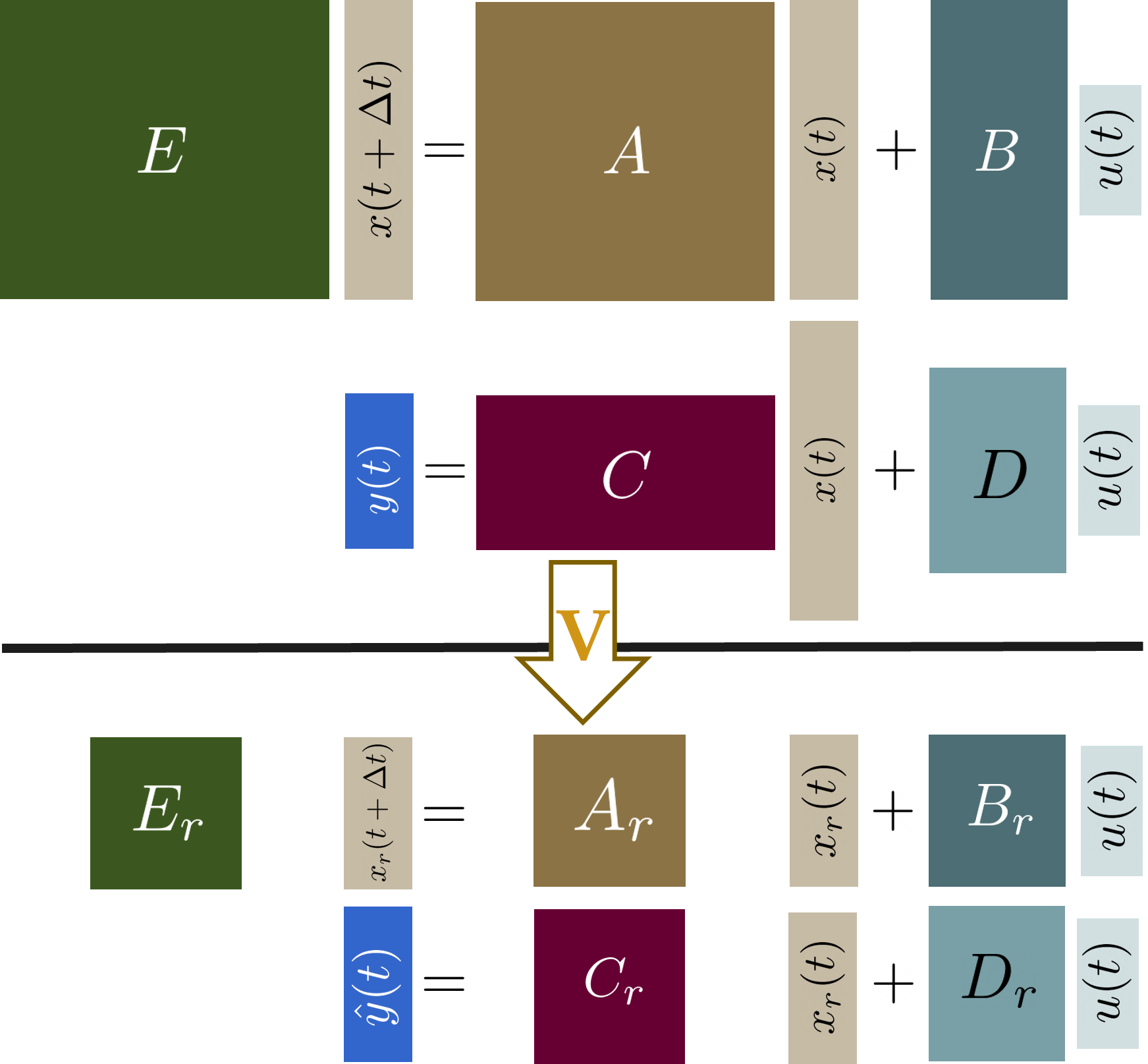}}{}\vspace{0cm} \hspace{0.1cm}
	\subfloat[\label{fig:MOR2}]{\includegraphics[keepaspectratio=true,scale=0.5]{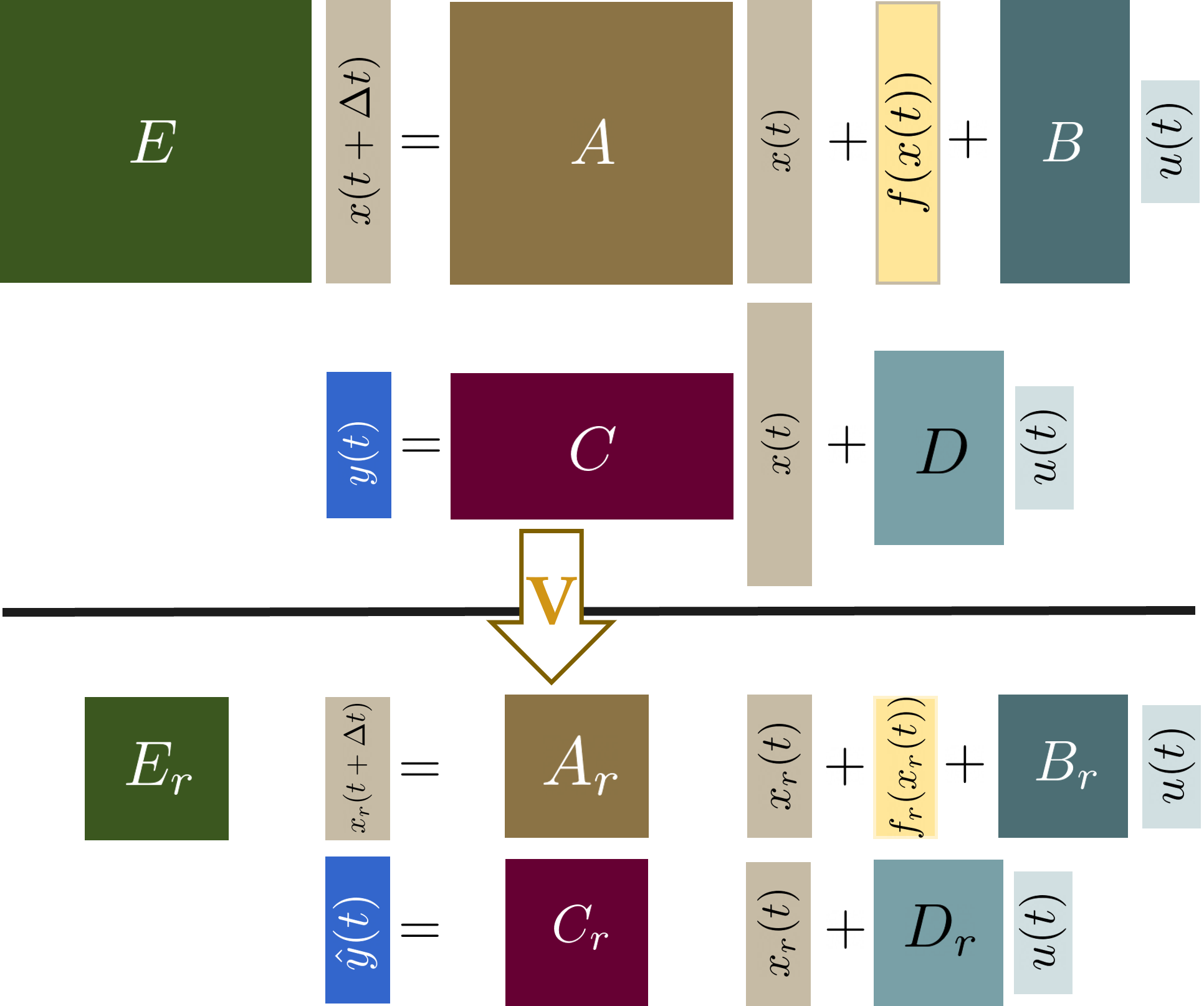}}{}\vspace{0cm}\hspace{0.1cm}
	\caption{(a) Linear and (b) nonlinear MOR methods configuration.}
	\label{fig:Net1Rest}
\end{figure}

\subsection{Model Order Reduction and Transformation Algorithms}
In our study, we investigate two SVD-based projection methods; POD and BPOD. The reason behind not applying BT method is that it has been proven to be computationally impractical for the linear water quality model \cite{wang2022model}. Both POD and BPOD are applied on the linearized MS-WQM, while an extension to the POD method is applied to reduce the nonlinear model where the nonlinear term is directly evaluated. 

%In advance, we propose a transformation model that can be counted a water quality oriented driven by the fact that the reason behind the high dimensionality of the system is the segmentation of each pipe, we refer to this method as water quality pipe segmentation transformation (WQ-PST).

Before explaining the detailed approach of the aforementioned methods, we start by explaining the general approach of SVD-based methods where a snapshot of the original space is taken. For the general nonlinear state-space representation \eqref{equ:NDE} which can concisely formulated as follows
\begin{equation}\label{eq:NL_SS}
	\begin{aligned}
		\mE(t)\vx(t+\Delta t) &= \mA(t) \vx(t) + \mB(t) \vu(t) + \vf(\vx(t)), \\
		\vy(t) &= \mC(t) \vx(t) + \mD(t) \vu(t),
	\end{aligned}
\end{equation}
the first step is to map the representation states $\vx \in \mathbb{R}^{n_x}$ to another space state $\vw \in \mathbb{R}^{n_x}$. This mapping aims to re-order the states according to their \textit{influence} in the preserved property. Driven by the goal of applying control algorithm on our model, we care to capture the most controllable and observable snapshots of the original space. Transformation is performed through constructing a non-singular matrix $\mV \in \mathbb{R}^{n_x \times n_x}$, so that $\vx=\mV \vw$. That is, Eq. \eqref{eq:NL_SS} is expressed in terms of $\vw$ as follows
\begin{equation}\label{eq:NL_SSw}
	\begin{aligned}
		\mE_w(t)\vw(t+\Delta t) &= \mA_w(t) \vw(t) + \mB_w(t) \vu(t) + \mV^{-1}\vf(\mV\vw(t)), \\
		\vy_w(t) &= \mC_w(t) \vw(t) + \mD(t) \vu(t),
	\end{aligned}
\end{equation}
where $\mE_w = \mV^{-1} \mE \mV, \; \mA_w = \mV^{-1} \mA \mV, \; \mB_w = \mV^{-1} \mB$ and $\mC_w = \mC \mV$.

Next, the reduced-order model is captured from the transformed mapping with number of states $n_r \ll n_x$ donated by $\vx_r \in \mathbb{R}^{n_r}$. A snapshot is taken of $\vx$ equal to $\mV_r \vx_r$ where $\mV_r$ is the matrix comprised of the first $n_r$ columns of $\mV$. Similarly, we define $\mL_r$ as the first $n_r$ rows of $\mV^{-1}$. Finally, the reduced-order model is expressed as
\begin{equation}\label{eq:NL_SSxr}
	\begin{aligned}
		\mE_r(t)\vx_r(t+\Delta t) &= \mA_r(t) \vx_r(t) + \mB_r(t) \vu(t) + \vf_r(\vx_r(t)), \\
		\overline{\vy}_r(t) &= \mC_r(t) \vx_r(t) + \mD(t) \vu(t),
	\end{aligned}
\end{equation}
where $\mE_r = \mL_r \mE \mV_r, \; \mA_r = \mL_r \mA \mV_r, \; \mB_r = \mL_r \mB$ and $\mC_r = \mC \mV_r$.

The choice of $n_r$ can be done arbitrarily as a fixed number or to conserve a specified level of energy between ROM and FOM. The energy of a system is determined by the summation of its eignvalues, hence $n_r$ can be chosen to keep a certain energy percentage of FOM in ROM \cite{lall1999empirical}. However, we investigate choosing different numbers of $n_r$ for each case study where the energy persevered is increased with larger $n_r$. 

Additionally, majority of MOR methods deal with original systems with zero initial conditions which does not align with the nature of water quality dynamics. Previously, the authors in \cite{wang2022model} have dealt with that by recognizing the non-zero initials network-wide as inputs for the system and setting $\hat{\vx}(t)=\vx(t)-\vx(0)$ in the original model. We follow same approach with further analysis for the nonlinear term of the mutual dynamics. The mutual reaction dynamics as stated in Section "\nameref{sec:MSmodel}" take place in pipes and tanks. That is, vector $\vf$ contains zeros except for states of pipes' segments and tanks. We define $\vx_{\mathrm{MS}_1}(t) := \{ c^\mathrm{TK}(t), \; c^{\mathrm{P}}(t)\}$ and $\vx_{\mathrm{MS}_2}(t) := \{ \tilde{c}^\mathrm{TK}(t), \; \tilde{c}^{\mathrm{P}}(t)\}$. Accordingly,
\begin{equation}~\label{eq:fMS}
	\vf(\vx_{\mathrm{MS}_1}(t),\vx_{\mathrm{MS}_2}(t)) = \m\alpha \boldsymbol{\cdot} \vx_{\mathrm{MS}_1}(t) \boldsymbol{\cdot} \vx_{\mathrm{MS}_2}(t), 
\end{equation} 
where $\m\alpha := \{\m\alpha_\mathrm{TK}, \; \m\alpha_\mathrm{P} \}$; $\alpha^\mathrm{TK}_j=-k_r \Delta t \frac{V_j^\mathrm{TK}(t)}{V_j^\mathrm{TK}(t+\Delta t)} \; \forall \; j=1,\dots, n_\mathrm{TK}$; and $\alpha_l^\mathrm{P} = -k_r \Delta t  \; \forall \; l=1,\dots, \sum_{i=1}^{n_\mathrm{P}} s_{L_i}$.

Henceforward, by setting $\hat{\vx}_{\mathrm{MS}_1}(t)={\vx}_{\mathrm{MS}_1}(t) - {\vx}_{\mathrm{MS}_1}(0)$ and $\hat{\vx}_{\mathrm{MS}_2}(t)={\vx}_{\mathrm{MS}_2}(t) - {\vx}_{\mathrm{MS}_2}(0)$ and substituting in \eqref{eq:fMS} we get

\begin{align}~\label{eq:fMSIn}
	&	\vf(\hat{\vx}_{\mathrm{MS}_1}(t),\hat{\vx}_{\mathrm{MS}_2}(t)) = \m\alpha \boldsymbol{\cdot} \hat{\vx}_{\mathrm{MS}_1}(t) \boldsymbol{\cdot} \hat{\vx}_{\mathrm{MS}_2}(t) \nonumber = \m\alpha \boldsymbol{\cdot} (\vx_{\mathrm{MS}_1}(t) - \vx_{\mathrm{MS}_1}(0)) \boldsymbol{\cdot} (\vx_{\mathrm{MS}_2}(t)-\vx_{\mathrm{MS}_2}(0)) \nonumber \\
	& = \m\alpha \boldsymbol{\cdot} (\vx_{\mathrm{MS}_1}(t) \boldsymbol{\cdot} \vx_{\mathrm{MS}_2}(t) - \textcolor{blue}{\vx_{\mathrm{MS}_2}(0) \boldsymbol{\cdot} \vx_{\mathrm{MS}_1}(t)} -	\textcolor{blue}{\vx_{\mathrm{MS}_1}(0) \boldsymbol{\cdot} \vx_{\mathrm{MS}_2}(t)} + \textcolor{indiagreen}{\vx_{\mathrm{MS}_1}(0) \boldsymbol{\cdot} \vx_{\mathrm{MS}_2}(0)}),
\end{align} 
which proves that considering $\vf(\hat{\vx}_{\mathrm{MS}_1}(t),\hat{\vx}_{\mathrm{MS}_2}(t))$ can be utilized by updating $\mA(t)$ in the original model to eliminate the negative terms (in blue) for pipes and tanks, while 
%add constant vector $\m\Psi$ to 
the positive constant term (in green) encapsulates the nonlinear term at the initial concentrations which is already considered (refer to the online version of the paper for the actual colors). Subsequently, the full-order model is formulated as
\begin{equation}\label{eq:NLSSIn}
	\begin{aligned}
		\mE(t)\hat{\vx}(t+\Delta t) &= \hat{\mA}(t) \hat{\vx}(t) + \hat{\mB}(t) \hat{\vu}(t) + \vf(\hat{\vx}(t)), \\
		\vy(t) &= {\mC}(t) \hat{\vx}(t) + \hat{\mD}(t) \hat{\vu}(t), 
	\end{aligned}
\end{equation}
where $\hat{\mA}(t)= \begingroup 
\setlength\arraycolsep{2pt}
\begin{bmatrix}
	{\mA}_{11}(t) & \hat{\mA}_{12}(t) \\ \hat{\mA}_{21}(t) & {\mA}_{22}(t)
\end{bmatrix}
\endgroup,$,  $ \hat{\mB}(t)=[\mB(t) \;\;\; \mA(t)\vx(0)],$  $\hat{\mD}(t)=[\mD(t) \;\;\; \mC(t)\vx(0)]$, and $\hat{\vu}(t) = [\vu^\top(t) \;\;\; \boldsymbol{1}^\top]^\top$.

On the other hand, for the linearized full-order model in \eqref{systemstatesLDE} same approach as in \cite{wang2022model} is followed and the final model formulated as

\begin{equation}\label{eq:LSSIn}
	\begin{aligned}
		\mE(t)\hat{\vx}(t+\Delta t) &= \hat{\mA}(t) \hat{\vx}(t) + \hat{\mB}(t) \hat{\vu}(t) + \m\Phi, \\
		\vy(t) &= {\mC}(t) \hat{\vx}(t) + \hat{\mD}(t) \hat{\vu}(t).
	\end{aligned}
\end{equation}
where $\hat{\mA}(t)= \breve{\mA}(t)$, $ \hat{\mB}(t)=[\mB(t) \;\;\; \breve{\mA}(t)\vx(0)],$  $\hat{\mD}(t)=[\mD(t) \;\;\; \mC(t)\vx(0)]$ and $\hat{\vu}(t) = [\vu^\top(t) \;\;\; \boldsymbol{1}^\top]^\top$.

Lastly, we judge the performance of the MOR methods by calculate the root-mean-square error (RMSE) metric, 
\begin{equation}\label{eq:RMSE}
	\mbox{RMSE}=\sqrt{\frac{1}{N_p}\sum_{j=1}^{N_p}||\vy(j)-\overline{\vy}(j)||^2_2}.
\end{equation}
The error is calculated for a specific simulation period of $N_p$ time-steps through which we apply same system inputs $\vu(j)$ to the two models. 

In the following sections, we give full description of the utilized methods. We start with applying POD and BPOD for the linearized formulation of the system, followed by integrating and handling the nonlinearity in the original representation of the system (Eq. \eqref{equ:NDE} for case of zero initial conditions and Eq. \eqref{eq:NLSSIn} for case of non-zero initial conditions).

The basic and the balanced POD methods are considered data-driven SVD methods. The main idea is to build empirical Gramians based on snapshots of the original system. These empirical Gramians avoid solving complicated, intractable in many case, Lyapunov equations. POD method relies on constructing controllability Gramian while BPOD constructs finite horizon controllability and observability Gramians. Notably, POD method favors highly controllable states over highly observable but less controllable ones which BPOD averts by reflecting observability in the captured snapshot. 

It is important to highlight that in our system the concepts of \textit{controllability} and \textit{observability} for the two chemicals are different in what they do reflect. While the input vector $\vu_1(t)$ depicts chlorine injections into the system by source or rechlorination stations, vector $\vu_2(t)$ enables simulating the intrusion of the contaminant to the system \cite{diao2013controllability}. Henceforward, controllability for the second chemical is indicating which network components get exposed/affected by the contamination event. On the other hand, typically water quality sensors are located to measure chlorine levels and from here comes the abstract concept of the system being observable for water quality measurements. This is a main reason for chlorine monitoring to be a solid proxy of the water quality state in a specific network. However, no sensors are placed for contaminants detection specifically with their wide range. That is, their observability is reflected in chlorine levels and not quantifiable in the matrix $\mC_{22}$ of \eqref{equ:NDE2} (i.e., a zero matrix). That puts a limitation on applying BPOD method as it will overlooks this contaminant because it is not observable. In Section "\nameref{sec:BPOD}", we propose a special approach to solve this issue. In addition, with no output measurement for that chemical, the RMSE metric in Eq. \ref{eq:RMSE} only measures the error for chlorine. In fact, the main purpose of this work is to control and monitor chlorine under contamination events which makes it valid to focus on the output of measuring its concentrations that are accurately representing the real-time state. Nevertheless, to evaluate the performance of the applied MOR methods we assume the existence of \textit{"imaginary"} sensors on some specific nodes to measure the fictitious reactant concentrations to calculate the corresponding error. 

In the following subsections, we explain what snapshots each method captures and how to construct these Gramians correspondingly. 

\subsubsection{Proper Orthogonal Decomposition (POD)}\label{sec:POD}

This method captures snapshot matrix $\mX_m$ that is built for specific number of steps $m$ by concatenating the states vector into 
\begin{equation}~\label{eq:Xm}
	\mX_m = [\vx(0) \;\; \vx(1) \;\; \dots \;\; \vx(m-1)],
\end{equation}
where $\mX \in \mathbb{R}^{n_x \times m}$.

The approximate $m$-step controllability Gramian $\mW_{C_m}$ is defined as $\mX_m \mX_m^\top \in \mathbb{R}^{n_x \times n_x}$. Next, we apply eigenvalue decomposition (ED) $\mW_{C_m}\mV = \mV \m\Lambda$ and obtain $\mV$ whose columns are the corresponding eigenvectors. However, in many cases applying ED for an $n_x \times n_x$ matrix with large $n_x$ is taxing. This can be avoided in cases of $m \ll n_x$ by constructing $\widetilde{\mW}_{C_m}=\mX_m^\top \mX_m \in \mathbb{R}^{m \times m}$. Accordingly, the eigenvalue decomposition procedure performing is easier and requires less computational time \cite{luchtenburg2011model}. In this case, ED is formulated as $\widetilde{\mW}_{C_m}\mQ = \mQ \m\Lambda$ where $\m\Lambda$ is the diagonal matrix of eigenvalues and matrix $\mQ$ is assembled with eigenvectors as columns. The transformation matrix is then calculated as $\mV=\mX_m \mQ \m\Lambda^{-\frac{1}{2}}$. For detailed step-by-step depiction of the POD method, follow Procedure \ref{alg:POD}. This procedure is followed for both chemicals.

{\setcounter{algocf}{0}
	\begin{algorithm}[t]
		\small	\DontPrintSemicolon
		\SetAlgorithmName{Procedure}
		1 1Construct snapshot $\mX_m$ as in \eqref{eq:Xm}\;
		\uIf{$n_x \ll m$}{Calculate $\mW_{C_m}=\mX_m \mX^\top_m$ \;
			Obtain transformation matrix $\mV$ by applying eigenvalue decomposition $\mW_{C_m}\mV = \mV \m\Lambda$}
		\Else{Calculate $\widetilde{\mW}_{C_m}=\mX_m^\top \mX_m$ \;
			Obtain matrices forms of eigenvector and eigenvalue of $\widetilde{\mW}_{C_m}$; $\mQ$ and $\m\Lambda$ \;
			Calculate transformation matrix as $\mV=\mX_m \mQ \m\Lambda^{-\frac{1}{2}}$ \;
		}
		Specify $n_r$ \;
		Define $\mV_r$ as the first $n_r$ columns of $\mV$\;
		Define $\mL_r$ as the first $n_r$ rows of $\mV^{-1}$\;
		Calculate $\mE_r, \mA_r, \mB_r, $ and $ \mC_r$ \;
		\uIf{FOM is nonlinear}{
			Follow Procedure \ref{alg:NNMOR}
		} 
		\textbf{end if}
		\caption{POD for general MS-WQM\label{alg:POD}}
\end{algorithm}}

\paragraph{Mapping and Integrating the Nonlinearity}\label{sec:NL-MOR}

While applying MOR, the reason behind separating the linear term(s) and the nonlinear term(s) is to be able to capture the behavior of the latter while working in a subspace of the original system (i.e., $\mathcal{R}^{n_r}$ instead of $\mathcal{R}^{n_x}$). In Eq. \eqref{eq:NL_SSxr}, following the projection of the whole system the nonlinear term is expressed as $\vf_r=\mL_r\vf(\mV_r \vx_r(t))$. Yet, the computational complexity of the nonlinear term still depends on $n_x$;

\begin{equation*}
	\vf_r=\underbrace{\mL_r}_{n_r \times \textcolor{red}{n_x}} \underbrace{\vf(\mV_r \vx_r(t))}_{\textcolor{red}{n_x} \times 1}.
\end{equation*}

Henceforward, it is proposed to reduce the nonlinear term based on an approximate hyperreduction approach. The approach is to measure not the full state-space variables, but particular points and from those points we construct the nonlinear term by interpolation around these points. In our study we specify the number of these points to equal $n_r$;

$$ \vf_r=\underbrace{\mL_r \mU_{f_r}}_{n_r \times \textcolor{blue}{n_r}} \underbrace{\hat{\vf}(t)}_{\textcolor{blue}{n_r} \times 1}.$$

The goal is to project $\vf(\mV_r \vx_r(t))$ onto $\mU_{f_r}$ so that $\vf(\mV_r \vx_r(t)) \approx \mU_{f_r} \hat{\vf}(t)$ and $\mL_r \mU_{f_r}$ can be pre-computed offline. This approach is called the \textit{"Gappy method"} of \textit{Galerkin projection} and the \textit{Discrete Empirical Interpolation Method (DEIM)} is used to reconstruct the nonlinear vector by interpolation. 
%That is, only a subset of the $n_r$ entries of the full system can be used as a case of a sparse Jacobian. 
We adopt a  \textit{Greedy sampling algorithm} to construct the measurement matrix to select the entries used. 

\begin{figure}[t!]
	\centering
	\hspace*{-0.3in}
	\includegraphics[width=0.6\textwidth]{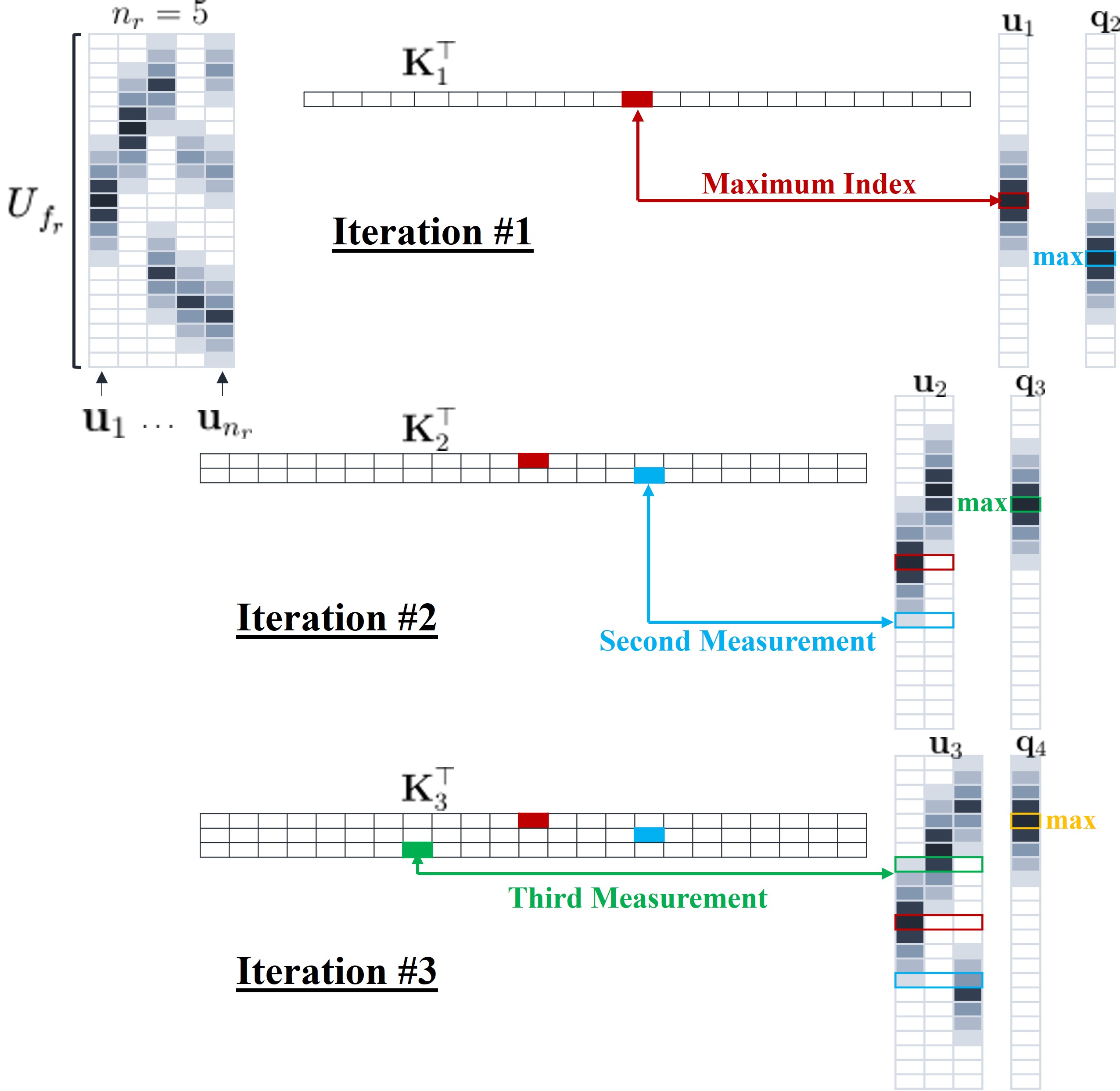}
	\caption{An illustrative example of applying the Greedy sampling algorithm to construct the measurement matrix $\mK$ for the case of $n_r=5$.}~\label{fig:DEIM}
\end{figure}

We start by stacking numerical snapshot $\mF_m$ only for the nonlinear term,

\begin{equation}~\label{eq:fsnapshot}
	\mF_m=[ \vf(\vx(0)) \;\; \vf(\vx(1)) \;\; \dots \;\; \vf(\vx(m-1)) ],
\end{equation}
followed by performing a \textit{separate} SVD for that snapshot, $\mF_m=\mU_f \m\Sigma_F \mQ_f^\top$. The next step is to define a rank-$n_r$ approximating basis $\mU_{f_r}$ as the first $n_r$ columns of $\mU_{f}$. Next, we construct the measurement matrix $\mK$ by applying the Greedy sampling algorithm as summarized in Procedure \ref{alg:NNMOR}. As shown in Fig. \ref{fig:DEIM}, the Greedy sampling algorithm starts by choosing the index with the maximum value in the first mode $\vu_1$ and making it the first measurement location. In the second iteration and the subsequent ones, we compute the residual to evaluate how the current measurement subspace projects onto the next one and decide on the next measurement point. The reason behind choosing the measurement with the maximum residual is that the modes are no longer orthogonal in the support space, hence, we calculate the residuals and locate the index with the maximum residual.

\begin{algorithm}[h!]
	\small	\DontPrintSemicolon
	\SetAlgorithmName{Procedure} 
	11 Capture $\mF_m$ as in \eqref{eq:fsnapshot} \;
	Perform SVD of $\mF_m=\mU_f \m\Sigma_F \mQ_f^\top$ \;
	Construct $\mU_{f_r}$ as the first $n_r$ columns of $\mU_{f}$ \;	
	\St{Greedy sampling algorithm for selecting the indices (entries of $\vf$)}{\KwIn {$\mU_{f_r}=[\vec{u}_{1} \; \dots \; \vec{u}_{n_r}]$}
		\KwOut{$\mathcal{I}:=\{i_1 \; \dots \; i_{n_r}\}$ and $\mK=[\ve_{i_1} \; \dots \;  \ve_{i_r}]$} 
		$[s,i_1] = \max \{|\vec{u}_1|\}$ \;
		$\mU_{f_r}=[\vec{u}_1], \; \mK=[\ve_{i_1}]$ \;
		\For{$I=2:n_r$}{solve $\mK^\top \mU_{f_r}\vb=\mK^\top \vec{u}_I$ for $\vb$ \;
			$\vq=\vec{u}_I - \mU_{f_r}\vb$ \;
			$[s,i_I]=  \max \{|\vq|\}$ \;
			$\mU_{f_r}=[\mU_{f_r}, \; \vec{u}_I], \; \mK=[\mK, \; \ve_{i_I}]$}
	}
	Calculate $\hat{\vf}(t)= (\mK^\top \mU_{f_r})^{-1}  \vf(\mK^\top\mV_r\vx_r(t))$
	\caption{Nonlinearity handling in MOR \label{alg:NNMOR}}
\end{algorithm}

\subsubsection{Balanced Proper Orthogonal Decomposition (BPOD)}\label{sec:BPOD}

The advance in the BPOD method is the reflection of both controllability and observability in ranking the states, unlike POD. This is attained by constructing two snapshots of the system, $\widetilde{\mX}_m$ which captures the impulse responses when applying impulse signal as system input (i.e., $u_i(m)=\gamma(m)$) and $\mP_m$ is assembled from states $\vp(t)$ obtained from the adjoint system with impulse response in the measurements as the system's output. For the linearized model in \eqref{eq:LSSIn}, the adjoint system can be expressed as follows,
\begin{equation}~\label{eq:LinAdj}
	\vp(t+\Delta t) = \breve{\mA}^\top(t) (\mE^{-1}(t))^\top  \vp(t) + \mC^\top(t) \vy(t) + \mE^{-1}(t){\m{\Phi}}.
\end{equation}
%An approximation to build the snapshots for the linearized system under steady hydraulic profiles is included in \ref{App:BPOD}.

%of the original system with $(\mA^\top,\mB^\top,\mC^\top,\mD^\top)$ with impulse response in the measurements in this case,
%\begin{equation}~\label{eq:XmYm}
%	\mX_m = [\vx(0) \; \dots \vx(m-1)], \; \mP_m = [\vp(0) \; \dots \vp(m-1)].
%\end{equation}

Next step is performing SVD to the block Hankel matrix $\mH_m=\mP_m^\top \widetilde{\mX}_m=\mU \m\Sigma \mQ^\top$ then specifying $n_r$ to collect the largest $n_r$ singular values in $\m\Sigma$ and obtain the corresponding left and right singular vectors (i.e., $\mU_r$ and $\mQ_r$). Accordingly, $\mV_r$ and $\mS_r$ are calculated as
\begin{equation}~\label{eq:BPODVrSr}
	\mV_r=\widetilde{\mX}_m\mQ_r\m\Sigma_r^{\frac{1}{2}}, \;\; \mL_r=\m\Sigma_r^{\frac{1}{2}}\mU_r^\top\mP_m^\top
\end{equation}

This approach is applicable for chlorine with sensors placed to measure its levels. For the fictitious reactant representing the contaminant, matrix $\mC_{22}$ in \eqref{equ:NDE2} is a zero matrix representing a non-sensed variables in our system. To solve this issue we assume that the contamination event is detected and the source location is determined. 
%Accordingly, contaminant levels is known at this location (i.e., observed).
This is considered a valid assumption in water quality monitoring to work backward detecting, classifying, and quantify using conventional WQ sensors \cite{yang2009real}. 
%This implies that for the fictitious reactant the intrusion location(s) are also observed which connotes that the controllable variables are observable at the same time.
This is different than the \textit{"imaginary"} sensors that are aforementioned while calculating the error to evaluate the performance of the applied methods. 

%It is noteworthy to declare that for a model simulating only the evolution of the fictitious reactant, the snapshots $\widetilde{\mX}_m$ and $\mP_m$ are identical and technically a representation of the snapshot $\mX_m$ captured for the POD method. This connotes that the controllable variables are observable at the same time.

Another advance of the BPOD is the ability of stabilizing it by choosing the length of the snapshots to be large enough to represent the actual Graminas shooting for infinity. We adopt an \textit{a priori stabilization} method to ensure that the snapshot captures the chemicals' evolution from the time it is injected in the system till it is observed by the furthest sensor. This is fulfilled by assembling the snapshots over a period exceeding $\underline{m}=\max \Big(\Big\lceil \frac{T_{BS}}{\Delta t}\Big\rceil \Big)=\max \Big(\Big\lceil \sum\frac{L^{BS}_i}{v^{BS}_i \Delta t}\Big\rceil \Big)$ where $L^{BS}_i$ and $v^{BS}_i$ are the length and velocity of the pipes the chemical travels through from a booster station to the furthest sensor. With the existence of multiple booster stations and sensors and within the simulation period, $\underline{m}$ is taken as the length corresponding to the maximum travel time $T_{BS}$. Accordingly, this method is affected by the actuators' and sensors' locations along the network.
%, more investigation regarding this effect included in Section \ref{sec:CaseStud}. 
Lastly, Procedure \ref{alg:BPOD} summarizes all the steps needed for a linear(ized) WQ model.

%\begin{algorithm}[h!]
%	\small	\DontPrintSemicolon
%	\SetAlgorithmName{Procedure}
%	1 1 \ForEach{Observable chemical}{	Construct snapshot $\widetilde{\mX}_m$ and $\mP_m$  \;
	%		Construct the block Hankel matrix $\mH_m=\mP_m^\top \widetilde{\mX}_m$ \;
	%		Perform SVD of $\mH_m=\mU \m\Sigma \mQ^\top$ \;
	%		Specify $n_r$ \;
	%		Obtain $\mU_r, \m\Sigma_r,$ and  $\mQ_r$ \;
	%		Calculate $\mV_r$ and $\mL_r$ via \eqref{eq:BPODVrSr} \;
	%		Calculate $\mE_r, \mA_r, \mB_r, $ and $ \mC_r$}
%	\ForEach{Non-observable chemical}{Follow Procedure \ref{alg:POD} till step 13}
%	\uIf{FOM includes nonlinear term}{
	%		Follow Procedure \ref{alg:NNMOR} to evaluate the nonlinear term
	%	} 
%	\textbf{end if}
%	\caption{BPOD for MS-WQM \label{alg:BPOD}}
%\end{algorithm}

\begin{algorithm}[h!]
	\small	\DontPrintSemicolon
	\SetAlgorithmName{Procedure}
	1 1 Obtain snapshots length $m=\underline{m}$ \;
	Construct snapshot $\widetilde{\mX}_m$ and $\mP_m$  \;
	Construct the block Hankel matrix $\mH_m=\mP_m^\top \widetilde{\mX}_m$ \;
	Perform SVD of $\mH_m=\mU \m\Sigma \mQ^\top$ \;
	Specify $n_r$ \;
	Obtain $\mU_r, \m\Sigma_r,$ and  $\mQ_r$ \;
	Calculate $\mV_r$ and $\mL_r$ via \eqref{eq:BPODVrSr} \;
	Calculate $\mE_r, \mA_r, \mB_r, $ and $ \mC_r$ \;
	%	\uIf{FOM is not linearized}{
		%		Follow Procedure \ref{alg:NNMOR}
		%	} 
	%	\textbf{end if}
	\caption{BPOD for linear(ized) WQM \label{alg:BPOD}}
\end{algorithm}

\section{Real-Time Regulation of MS-WQM via Model Predictive Control and McCormick Relaxations}~\label{sec:MS-MPC}
The water quality control problem is formulated over simulation period $[0,T_s]$ and constrained by putting standard upper and lower bounds on chlorine concentrations stated by EPA regulations \cite{acrylamide2national}, which are $x_1^{\min}=0.2$ mg/L and $x_1^{\max}=4$ mg/L. We note that the contaminant in the system is assumed to be detected and classified. Accordingly, for some toxic or health threatening substances a constraint can be introduced to be kept lower than the allowed concentration defined by EPA. These bound for both chemicals formulates the constraint $\vx^{\min} \leq \vx(t) \leq \vx^{\max}$. Additionally, the control inputs for chlorine are constrained to be non-negative and limited by the chlorine availability and capacity of each booster station. The objective of this control problem is to keep chemicals concentrations in all network's components within the aforementioned bounds while minimizing the cost of chlorine injections. That being said, the problem formulation is as follows,

\begin{subequations}~\label{eq:opt}
	\begin{align}
		\underset{\vx(t),\vu_1(t)}{\mbox{minimize}} \hspace{1cm} & \mathcal{J}(\vu_1(t)) = \mu \sum_{t=1}^{N_p} \vq^\mathrm{B}(t)^\top \vu_1(t) \\
		\begin{split}~\label{eq:optConst}
			\mbox{subject to} \hspace{1cm} & \mbox{WQM} \; \eqref{equ:NDE}, \\ & \vx^{\min} \leq \vx(t) \leq \vx^{\max}, \\ & \vu_1^{\min} \leq \vu_1(t) \leq \vu_1^{\max},
		\end{split}	
	\end{align}		
\end{subequations}

where problem variables $\vx(t)$ and $\vu_1(t)$ are chemicals concentrations network-wide and chlorine injections through booster stations, $\vq^\mathrm{B}(t)$ is the flow rates at the nodes corresponding to the locations of the booster stations, $\mu$ is the unit cost of chlorine in \$/mg, and WQM is the water quality model we are simulating and controlling following the representation in \eqref{equ:NDE}. Finally, $N_p$ is the number of time-step in the simulation period, $N_p = \frac{T_p}{\Delta t}$.
%(i.e., multi-species FOM, multi-species ROM, single-species FOM).

Nonetheless, this problem has large number of variables $\vx(t)$ and $\vu_1(t)$. This issue can be solved by transforming a constrained LP \eqref{eq:opt} to a quadratic program (QP) with fewer variables by applying real-time constrained model predictive control (WQ-MPC). The water quality control formulated in \cite{wang2021effective} is based on the linear state-space representation of the single-species WQ dynamics. In addition, same control algorithm is applied in \cite{wang2022model} for the reduced order model of the single-species representation and it proved its validity and effectiveness. For brevity, we do not include the details and the derivation in this paper for the case of linearized MS-WQM. Eventually, the WQC problem is formulated as quadratic program.
For the nonlinear MS-WQM, the nonlinearity in the constraint can be relaxed using \textit{McCormick envelopes} and integrated back to the original constrained control problem as explained in the following section.

\begin{figure}[t!]
	\centering
	\subfloat[\label{fig:MPC}]{\includegraphics[keepaspectratio=true,scale=0.45]{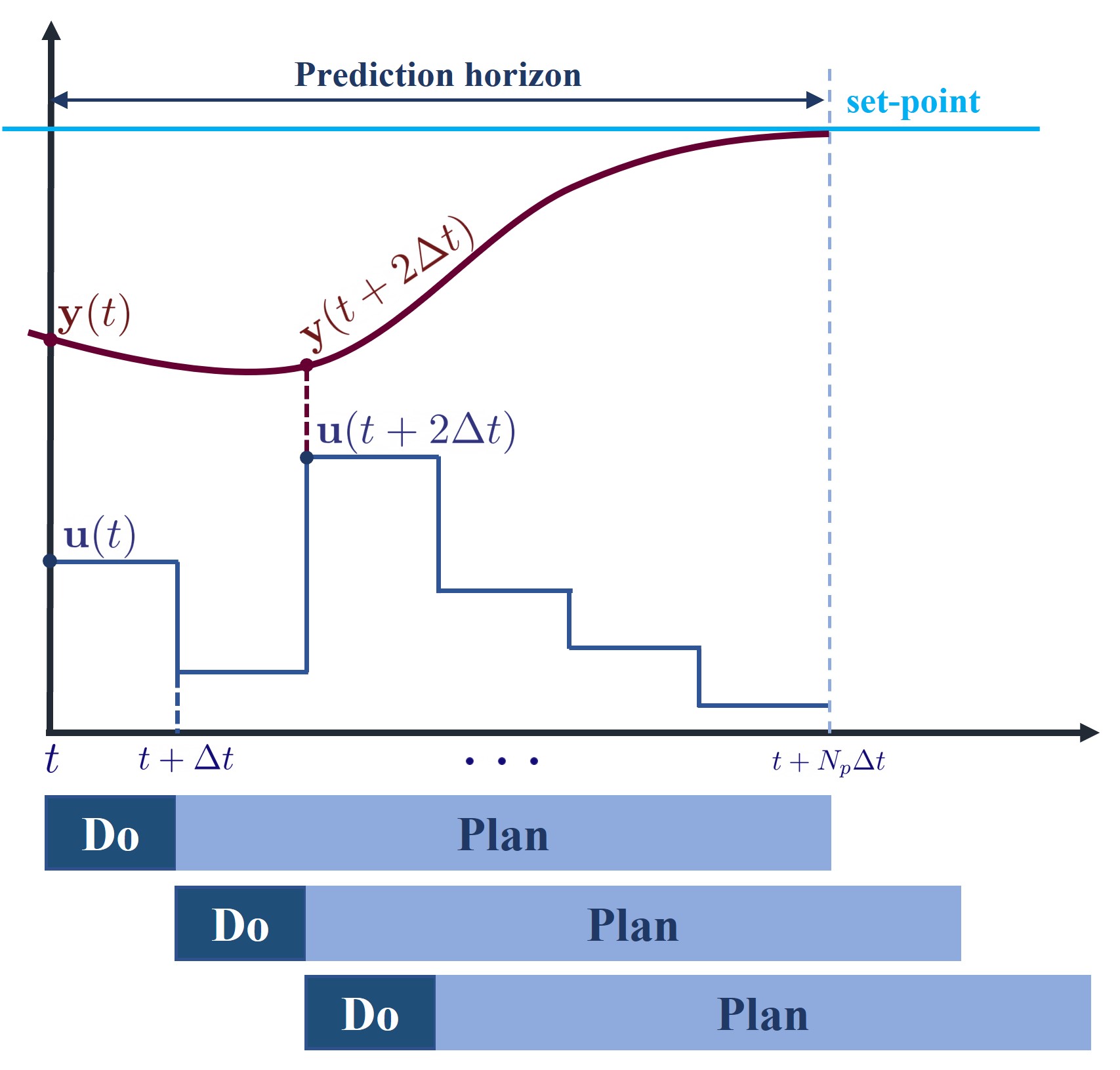}}{}\vspace{-0.05cm} \hspace{0.1cm}
	\subfloat[\label{fig:McC}]{\includegraphics[keepaspectratio=true,scale=0.4]{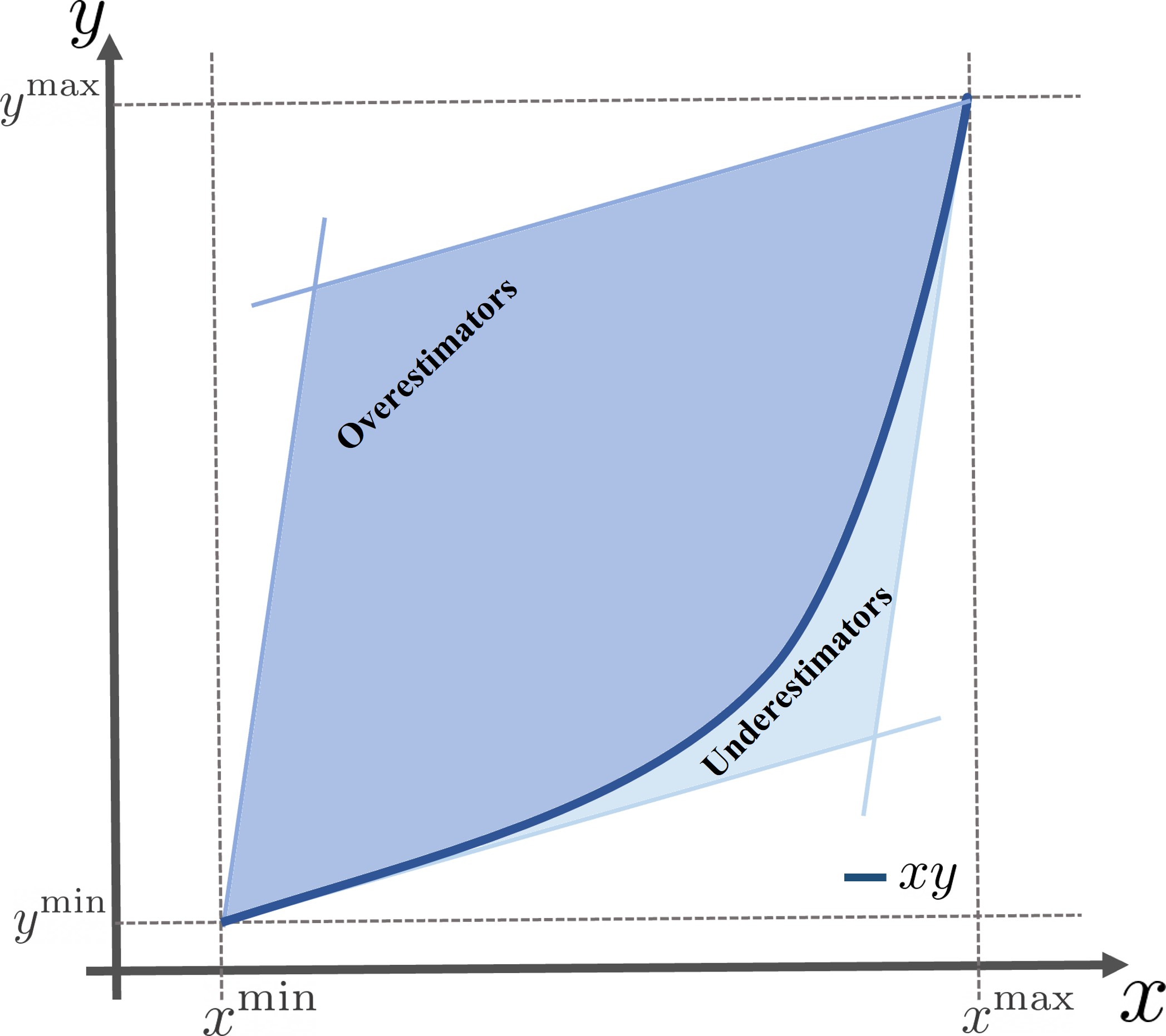}}{}\vspace{-0.05cm} \hspace{-0.2cm}
	\caption{(a) Discrete MPC prediction horizon scheme, and (b) graphical representation of McCormick envelope relaxation.}~\label{fig:MPC_McC}
\end{figure}

\subsection{McCormick Relaxation}~\label{sec:McC}
%\begin{wrapfigure}{r}{4cm}
%	\centering
%	\includegraphics[width=4cm]{McC}
%	\caption{Graphical representation of McCormick envelope relaxation.}~\label{fig:McC}
%\end{wrapfigure} 
The nonlinear term in the constraint is formulated as a bi-linear expression depending on the concentration of two chemicals at a specific network component. This constraint can be relaxed using \textit{McCormick Relaxation} for bilinear nonlinear problems \cite{mccormick1976computability}. This method turns the bilinear term into two envelopes surrounded by overestimators and underestimators to work within, Fig. \ref{fig:McC}. For a bilinear expression $z=x_1 x_2$ where $x_1$ and $x_2$ are the two chemicals concentrations under $x^{\min}_1 \leq x_1 \leq x^{\max}_1$ and $x^{\min}_2 \leq x_2 \leq x^{\max}_2$, $z$ is introduced as a new decision variable with the following constraints,
\begin{equation}\label{eq:McC}
	\begin{aligned}
		z \geq x^{\min}_1 x_2 + x_1 x^{\min}_2 - x^{\min}_1 x^{\min}_2, \\
		z \geq x^{\max}_1 x_2 + x_1 x^{\max}_2 - x^{\max}_1 x^{\max}_2, \\
		z \leq x^{\max}_1 x_2 + x_1 x^{\min}_2 - x^{\max}_1 x^{\min}_2, \\
		z \leq x^{\min}_1 x_2 + x_1 x^{\max}_2 - x^{\min}_1 x^{\max}_2.
	\end{aligned}
\end{equation}

We note that, in some cases the upper bound on $x_2$ is not specified or its concentration initially is lower than the maximum allowed one stated by EPA. In such cases, we specify $x_2^{\max}$ to be equal to this initial concentration detected to be able to tighten the overestimators envelope while having the minimum equal to zero.  

Eventually, the problem formulation explained for the linearized model can be adopted with these modifications. First, a new variable vector $\vz(t)$ is introduced and it replaces $\vf(\vx_1,\vx_2,t)$ in \eqref{equ:NDE1}. Additionally, the total number of the constraints added to the optimization problem via \eqref{eq:McC} is equal to $4(n_\mathrm{TK}+\sum_{i=1}^{n_\mathrm{P}} s_{L_i})$ as the nonlinear term is defined for pipes' segments and tanks and is the same for both chemicals at same element of the aforementioned (refer to Eq. \eqref{equ:MR_PTK}). To that end, the WQC problem described in \eqref{eq:opt} is modified as follows,
\begin{subequations}~\label{eq:optMcC}
	\begin{align}
		\underset{\vx(t),\vu_1(t), \vz(t)}{\mbox{minimize}} \hspace{1cm} & \mathcal{J}(\vu_1(t)) = \mu \sum_{t=1}^{N_p} \vq^\mathrm{B}(t)^\top \vu_1(t) \\
		\begin{split}
			\mbox{subject to} \hspace{1cm} & \mbox{WQM} \; \eqref{equ:NDE}, \\ & \vx^{\min} \leq \vx(t) \leq \vx^{\max}, \\ & \vu_1^{\min} \leq \vu_1(t) \leq \vu_1^{\max}, \\ & \mathrm{McCormick} \; \eqref{eq:McC}
		\end{split}	
	\end{align}		
\end{subequations}

%In Appendix \ref{App:McCMPC}, we transform \eqref{eq:optMcC} into a linear augmented formulation based on which the final WQC-QP is built.  It is worth to highlight that under this relaxation, control variables are increased to be $N_p n_{u_1} + n_x$. 

Next step is transforming \eqref{eq:optMcC} into a linear augmented formulation based on which the final WQC-QP is built. First, by introducing $\vz(t)$ into \eqref{equ:NDE1}, the state-space representation is updates as

\begin{equation}~\label{eq:NDEMcC}
	\vx(t+1) = \mA(t) \vx(t) + \mB(t) \vu(t) + \beta \vz(t).
\end{equation}
where $\beta=-k_r$. Then, we define the change in the states and inputs as follows

\begin{equation}
	\Delta \vx(t+1) = \vx(t+1) - \vx(t), \;\; \Delta \vu(t+1) = \vu(t+1) - \vu(t), \;\; \Delta \vz(t+1) = \vz(t+1) - \vz(t). 
\end{equation}

To concatenate these rates of change in \eqref{eq:NDEMcC}, $\Delta \vz$ is assembled to the vector of systems decision inputs to be optimally chosen within the envelopes defined by \eqref{eq:McC}. Eventually, we reach an the augmented state-space representation in \eqref{eq:McCMPC}.

\begin{equation}~\label{eq:McCMPC}
	\underbrace{\begin{bmatrix}
			\Delta \vx(t+1) \\ \vy(t+1) 
	\end{bmatrix}}_{\vx_a(t+1)} = \underbrace{\begin{bmatrix}
			\mA(t)  & \boldsymbol{0} \\
			\mC(t) \mA(t) & \mI
	\end{bmatrix}}_{{\m\Phi}_a}	\underbrace{\begin{bmatrix}
			\Delta \vx(t) \\ \vy(t) 
	\end{bmatrix}}_{\vx_a(t)}  + \underbrace{\begin{bmatrix}
			\mB(t) & \beta \\
			\mC(t) \mB(t) & \beta \mC(t) 
	\end{bmatrix}}_{{\mat{\Gamma}}_a}	 \underbrace{\begin{bmatrix}
			\Delta \vu(t) \\ \Delta \vz(t)   
	\end{bmatrix}}_{\Delta\vu_a(t)}
\end{equation}
This augmented representation can be abstractly rewritten as, $\vx_a(t+1)={\m\Phi}_a \vx_a(t+1)+{\mat{\Gamma}}_a \Delta\vu_a(t)$. To avoid redundancy, integrating this equality into WQC-MPC formation follows the same approach of \cite{wang2022model} reaching the final QP \cite[Eq. (38)]{wang2022model}. On another note, the added constraints expressed in \eqref{eq:McC} are incorporated in the constraints on the optimization variables. 

%	the system states $\vx \in \mathbb{R}^{n_x}$ is updated to be $\vx_z=[\vx \;\;\;  \vz] \in \mathbb{R}^{2n_x}$ for each time-step. Subsequently, matrix $\mA$ is updated as well to $\mA_z$ where }

%\begin{figure}[h!]
%	\centering
%	\includegraphics[width=0.28\textwidth]{McC}
%	\caption{Graphical representation of McCormick envelope relaxation.}~\label{fig:McC}
%\end{figure}

\subsection{Generalized Comprehensive Water Quality Modeling and Control Framework}~\label{sec:FW}
In our study, we have covered model order reduction and control for multi-species water quality dynamics where chlorine is reacting with another source of contamination in form of a bi-linear expression---refer to Section "\nameref{sec:MSmodel}". 
%As mentioned, it has been also covered for single-species model in \cite{wang2022model}. 
%However, the change in hydraulics which leads to change in states count has not been investigated. That being said, we apply MOR for the single-species WQ model with different hydraulic settings to showcase  
However, there are other formulations for single-species and multi-species chlorine bulk decay and reaction dynamics as listed in \cite{ELSHERIF2022}. We include a short list of these formulations in Tab. \ref{tab:CLModels}, nevertheless for more details and descriptions refer to the aforementioned study. The following generalized framework described in Algorithm 1 maps out the methods adopted in this study to be applied on the different decay and reaction models.

%\begin{table}[h!]
%	\centering
%		\caption{Chlorine bulk decay and reaction models expressions}~\label{tab:CLModels} 
%	\includegraphics[width=\textwidth]{TabToFig}
%\end{table}

\begin{table}[h!] 
\centering
\caption{Chlorine bulk decay and reaction models expressions}~\label{tab:CLModels} 
\setlength{\tabcolsep}{1.5pt}
{\begin{threeparttable}
		\begin{tabular}{c | L | c | c | c }
			\hline
			M\# & Model  & Model formulation & \#States & L/NL\tnote{a} \\
			\hline
			{M-1} & First-order & $\frac{dc}{dt}=-kc(t)$ & $n_x$ & L \\
			\hline
			{M-2} & First-order with stable component & $\frac{dc}{dt}=-k(c(t)-c_\mathrm{L})$ & $n_x$ & L \\
			\hline
			{M-3} & Parallel first-order & 
			\smaller{$\begin{aligned}
					& \frac{dc_1}{dt} \Big\vert_{\text{fast}}=-k_{\text{fast}}c_1(t) \\
					& \frac{dc_2}{dt} \Big\vert_{\text{slow}}=-k_{\text{slow}}c_2(t) \\
					& c_t(t) = c_1(t) + c_2(t)
				\end{aligned}$} & $2n_x$ & L \\
			\hline
			{M-4} & Parallel second-order & 
			\smaller{$\begin{aligned}
					& \frac{dc_\mathrm{F}}{dt} \Big\vert_{\text{fast}}=-k_{\text{fast}}c(t)c_\mathrm{F}(t) \\
					& \frac{dc_\mathrm{S}}{dt} \Big\vert_{\text{slow}}=-k_{\text{slow}}c(t)c_\mathrm{S}(t) \\
					& \frac{dc}{dt} = \frac{dc_\mathrm{F}}{dt} + \frac{dc_\mathrm{S}}{dt}
				\end{aligned}$} 	& $2n_x$	& NL			 \\
			\hline
			{M-5} & n\textsuperscript{th}-order & $\frac{dc}{dt}=-kc^n(t)$ & $n_x$ & NL \\
			\hline
			{M-6} & n\textsuperscript{th}-order with stable component & $\frac{dc}{dt}=-k(c(t)-c_\mathrm{L}) c^{(n-1)}$ & $n_x$ & NL \\
			\hline
			{M-7} & Second-order with fictitious component & \smaller{$\begin{aligned}
					& \frac{dc}{dt}=-k c(t) \tilde{c}(t) \\
					& \frac{d \tilde{c}}{dt}=-kc(t) \tilde{c}(t)
				\end{aligned}$} & $2n_x$ & NL \\
			\hline
			{M-8} & Second-order with multiple components & \smaller{$\begin{aligned}
					& \frac{dc_i}{dt}=-k_i c(t) \tilde{c}_i(t) \\
					& \frac{d \tilde{c}_i}{dt}=-k_i c(t) \tilde{c}_i(t)\\
					& \frac{dc}{dt}= \sum_{i}^I \frac{dc_i}{dt}
				\end{aligned}$} & $In_x$ & NL \\
			\hline
			\hline
		\end{tabular}
		\begin{tablenotes}
			\smaller {\item[a] L: Linear or NL: Nonlinear model expression.}
		\end{tablenotes}
\end{threeparttable}}
\end{table}

For the first-order, first-order with stable component, and  parallel first-order (M-1\&M-2\&M-3) models, the dynamics are linear and accordingly follow the procedure of the linearized model represented in our study. Whilst, the second-order with multiple components (M-8) is considered to be the same formula as the second-order with fictitious component (M-7) we cover in this paper except for the number of states which gets multiplied by the number of reactants in the system. That is, model order reduction for M-8 model becomes more demanded. On the other hand, the parallel second-order model (M-4) is a special form of the second-order with fictitious component. Lastly, the n\textsuperscript{th}-order without and with stable component models are higher order models which can be reduced as a nonlinear models or be transferred into quadratic approximation and apply piecewise linear relaxation.

%{\vspace{0.4cm}\hspace{-0.6cm}\noindent\includegraphics[width=\textwidth]{AlgToFig}}

\setcounter{algocf}{0}
\begin{algorithm}[h]
\small	\DontPrintSemicolon
\caption{Generalized water quality modeling and control framework \label{alg:FramWQ}}
\KwIn{WDN topology, components’ characteristics, and hydraulics
	parameters}
\KwOut{Real-time water quality states $\vx(t)$ and control inputs $\vu(t)$ at time $t$ of a simulation period of $T_s$} 
\Init{}{Define $\Delta t$, number of segments $s_i$ for each pipe and accordingly $n_x$ \;
	Formulate WQ state-space representation \eqref{equ:NDE} as explained in Section "\hyperref[sec:WQModel]{State-space Multi-species Water Quality Model}" and according to the reaction dynamics in Tab. \ref{tab:CLModels}.}
\uIf{Applying {M-1}/{M-2}/{M-3} reaction model}{
	Follow Procedure \ref{alg:BPOD} to obtain ROM \;
	Apply constrained real-time WQ-MPC on \eqref{eq:opt}}
\uElseIf{Applying {M-4}/{M-7}/{M-8} reaction model}{
	\uIf{Following Procedure \ref{alg:POD}} {Apply McCormick relaxation via \eqref{eq:McC} \;
		Apply constrained real-time WQ-MPC on \eqref{eq:optMcC}}
	\Else{\textit{Linearizing and following Procedure \ref{alg:BPOD}} \textbf{then} \;
		Apply constrained real-time WQ-MPC on \eqref{eq:opt}}}
\Else{\textit{Applying {M-5}/{M-6}} \textbf{then} \;
	Follow Procedure \ref{alg:POD} to obtain ROM \;
	Transforming into quadratic app./Apply piecewise linear relaxation \;
	Apply constrained real-time WQ-MPC on \eqref{eq:opt}
}
\end{algorithm}

To recapitulate, this paper is an extension of our previous work in \cite{ELSHERIF2022,wang2022model}. That is, some methods/aspects included in this section have already been covered in these studies. Yet, we have decided to reintroduce this material into our study in a more concise way to ensure that the reader can effectively follow and comprehend the new information being presented. In the following bullet points, we highlight the novelty in our work in comparison to these studies.
\begin{itemize}
\item Adopting a nonlinear multi-species water quality dynamics rather than the linear single-species dynamics in a control framework. These dynamics enable a more heightened level of realism in the system dynamics representation. 
\item Following two different paths where different MOR methods and control algorithms are applied on the original nonlinear and a linearized forms of the model. For the linearized model, we implement the same MOR techniques (specifically, POD and BPOD) described in the paper by \cite{wang2022model}. On the other hand, for the nonlinear model, we introduce the Gappy method, which employs a greedy algorithm to effectively handle the nonlinearity and reduce the model dimension.
\item Expanding the implementation of these MOR techniques for the case of non-zero initial conditions by developing a closed formulation that preserves the original nonlinear formulation of the model.
\item Likewise, for the linearized model we implement the MPC algorithm explain in \cite{wang2022model} to control and regulate chlorine levels under the multi-species dynamics. In contrast, we extend the MPC algorithm to incorporate the McCormick Relaxation technique, which is specifically tailored for the nonlinear model.
\item While the same methods and algorithms are employed for the linearized model as in the linear single-species model described in the paper by \cite{wang2022model}, special consideration is required when implementing these techniques for the linearized model. This is primarily due to the duplication of state numbers and the distinct construction of representation matrices for the fictitious reactant. These factors necessitate a specific approach to ensure accurate and reliable results during the implementation of these methods in the context of the linearized model.
\item Investigating and evaluating using an explicit vs. implicit discretization schemes from a control-theoretic standpoint. Specifically, we apply Upwind schemes, which offer a more accurate representation of the advection-reaction 1-D partial differential equations (PDEs) compared to the Lax-Wendroff scheme utilized in the work by \cite{wang2022model}. The advance in implementing an Upwind scheme is proved and demonstrated in \cite{ELSHERIF2022}. In addition, Lax-Wendroff scheme is an explicit scheme and the prior study does not extensively investigate the use of an implicit scheme in this context.
\item A novel component of our study is investigating the water quality control framework performance under different system hydraulic settings. These settings directly impact the water quality dynamics and their progression over time within the same network. This exploration adds a unique dimension to our study, shedding light on the interplay between system hydraulics and water quality dynamics for enhanced understanding and improved control strategies.
\end{itemize} 
The validity of these techniques and the performance of the framework are evaluated and substantiated in the subsequent section through a series of numerical case studies.  

\begin{figure}[h!]
\centering
\includegraphics[width=0.6\textwidth]{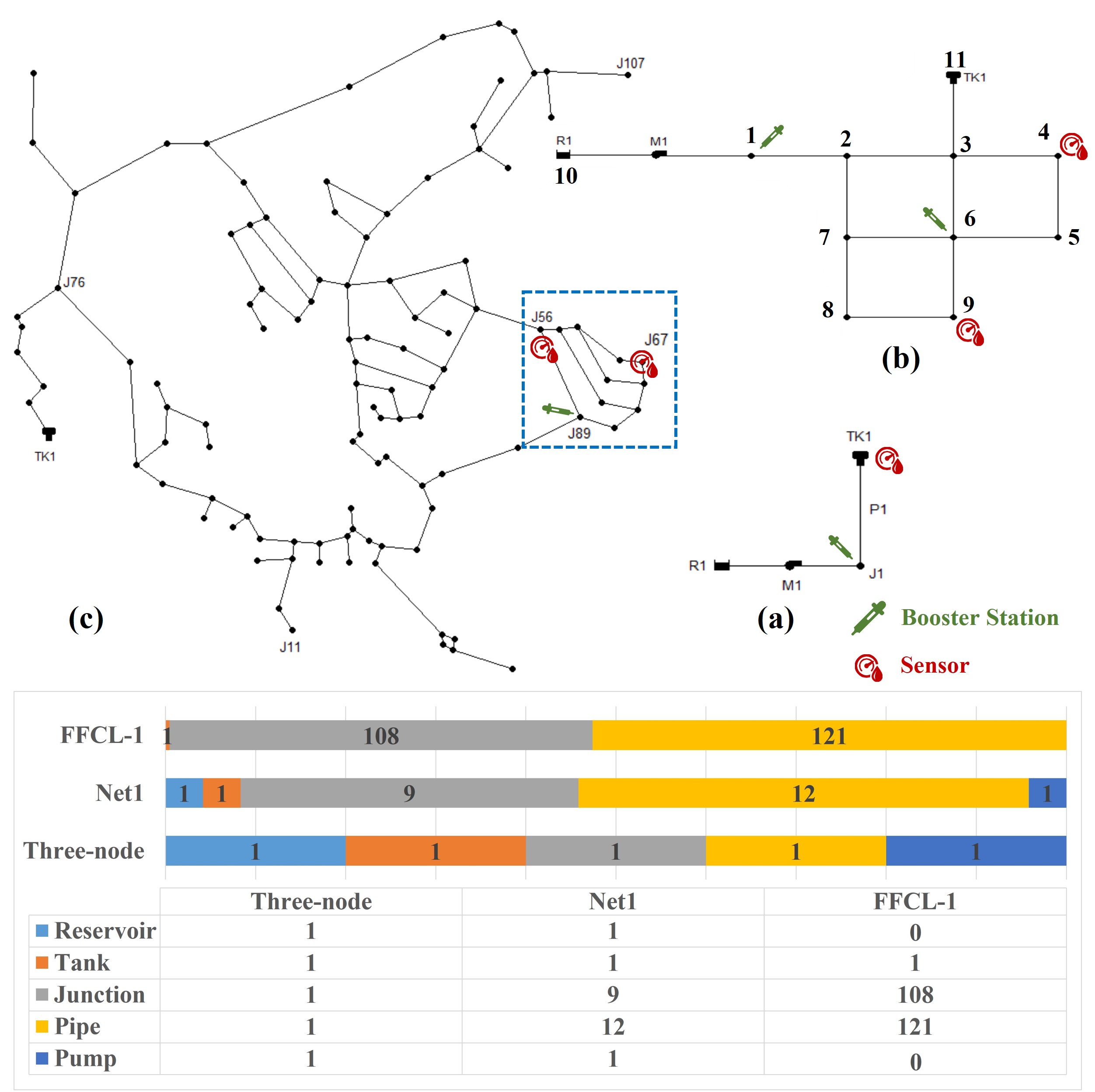}
\caption{Case studies’ layouts and their components count: (a) Three-node network, (b) Net1, and (c) FFCL-1 with the zone we control framed.}~\label{fig:CaseStudy}
\end{figure}

%\begin{figure}[h!]
%\centering
%\includegraphics[width=\textwidth]{Fig7.jpg}
%\caption{Test networks' components count.}~\label{fig:NetComp}
%\end{figure}

\section{Case Studies}~\label{sec:CaseStud}
This section demonstrates the proposed framework for model order reduction and control of MS-WQM. Particularly, we attempt to answer the following questions:
{\begin{itemize}[leftmargin=.3in]
	\item[$\triangleright$] \textit{Q1:} How does the number of operating points impact agreement between the linearized model and the nonlinear MS-WQM?    
	\item[$\triangleright$] \textit{Q2:} How effective are the proposed MOR producers in terms of accuracy and computational time when applied on the MS-WQM? 
	\item[$\triangleright$] \textit{Q3:} How sensitive is the performance of MOR and control algorithm to the discretization methods and system's hydraulics? 
	\item[$\triangleright$] \textit{Q4:} How reliable and robust is model predictive control when applied to control chlorine levels under multi-species dynamics?
\end{itemize}}

Numerical studies in this section are performed on three different networks, three-node, Net1, and FFCL-1 networks \cite{rossman2020epanet}. As shown in Fig. \ref{fig:CaseStudy}, each of the networks has different topologies and scales. The three-node network is a self-designed network to help provide simple illustrations for different approaches throughout our framework implementation. Net1 includes different types of network components and has a looped layout. The FFCL-1 network is based on the Fairfield, CA, USA water distribution system on which we test the scalability of our framework and its performance with scattered dead-ends. Also, Fig. \ref{fig:CaseStudy} illustrates and lists each of the networks' components. 

In addition to the listed components for each of the test networks in Fig \ref{fig:CaseStudy}, each network has a different number of sensors and booster stations. The three-node network has one booster station at Junction J1 and one sensor at Tank TK1. Net1 has two stations at Junctions 1 and 6 and sensors at Junctions J4 and J9. Lastly, the controlled region of the FFCL-1 network has two sensors at Junctions J56 and J67, and one rechlorination station at J89. 

It is worth mentioning that for any WDN, the system dimension depends on the hydraulics parameters and water quality simulation time-step which accordingly define the number of segments for each pipe (i.e., pipes state variables). Further, changing the velocities and flows from one scenario to another results in distinct chemical concentrations across the network components for each scenario. With that in mind, in some of our case studies, we feature the effect of changing the hydraulics for the same network. In some other case studies, we fix the hydraulics setting in the system to investigate/test a technique or an approach under discussion. In addition, for all studies performed in this section we use the Implicit Upwind scheme except for Section "\nameref{sec:ImpExpProf}" where we compare its performance with the Explicit Upwind scheme from a control-theoretic perspective.

%\begin{table}[h!]
%	\centering
%	\caption{Test networks components}~\label{tab:NetComp}
%	{\small\begin{tabular}{c|c|c|c|c|c}
	%			\hline
	%			\multirow{2}{*}{Network} & \multicolumn{5}{c}{Components Count} \\
	%			\cline{2-6}
	%			& Reservoir & Tank & Junction & Pipe & Pump \\
	%			\hline
	%			Three-node & 1 & 1 & 1 & 1 & 1 \\
	%			\hline 
	%			Net1 & 1 & 1 & 9 & 12 & 1  \\
	%			\hline 
	%			FFCL-1 & 0 & 1 & 108 & 121 & 0 \\ 
	%			\hline
	%			\hline
	%	\end{tabular}}
%\end{table}

\subsection{Nonlinear vs. Linearized Models}~\label{sec:CaseStudLin}
Studies \cite{chen1999model,schilders2008,liu2016control} state that applying a linear MOR algorithm on a linearized system gives satisfactory performance when the linearized system is close to the original nonlinear one or operating within/near its linear regime. In these studies, linearization is performed around one operating point for the whole simulation horizon. We apply the same approach by linearizing around two operating points, (0,0) and (0.2,0.05) mg/L for chlorine and fictitious reactant respectively at Tank TK1 of the three-node network. In this scenario, a constant demand is drawn from J1 and sources of 2 mg/L of chlorine and 0.5 mg/L of the fictitious reactant are provided at R1 and zero initial conditions for other network components. As demonstrated in Fig. \ref{fig:3NLin1}, linearization around the operating point of (0,0) results in higher concentrations compared to the nonlinear model (based on the NDE \eqref{equ:NDE}) for both chemicals due to the fact it drops out the nonlinear term and neglects the mutual reaction. On the other hand, linearizing the model around one random operating point as (0.2,0.05) mg/L results in relatively closer values for chlorine concentrations but not as close for the fictitious reactant. Furthermore, unlike this scenario, in real-time water networks hydraulics are not fixed and demands are time-variant resulting in chemical evolution with different schemes for which fixing the operating point for all elements is not actively efficient. That is, we investigate next taking different operating points for each network component along the simulation window every specific number of time-steps. 

\begin{figure}[t!]
\centering
\subfloat[\label{fig:3NLin1a}]{\includegraphics[keepaspectratio=true,scale=0.55]{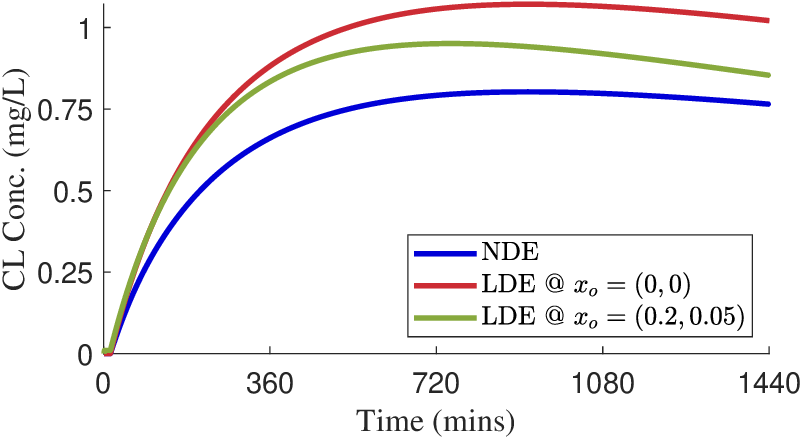}}{}\vspace{-0.05cm} \hspace{0.25cm}
\subfloat[\label{fig:3NLin1b}]{\includegraphics[keepaspectratio=true,scale=0.55]{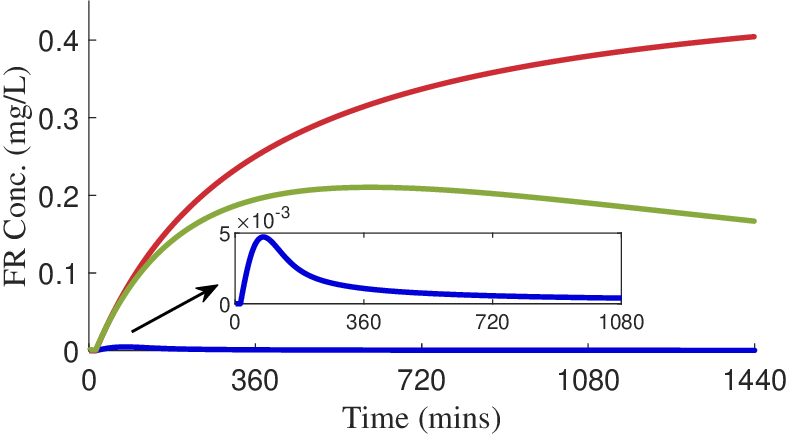}}{}\vspace{-0.05cm} \hspace{-0.1cm}
\caption{Nonlinear vs linearized models results for (a) chlorine and (b) fictitious reactant at TK1 of the three-node network.}
\label{fig:3NLin1}
\end{figure} 

\begin{figure}[h!]
\centering
\subfloat[\label{fig:3NLin2a}]{\includegraphics[keepaspectratio=true,scale=0.5]{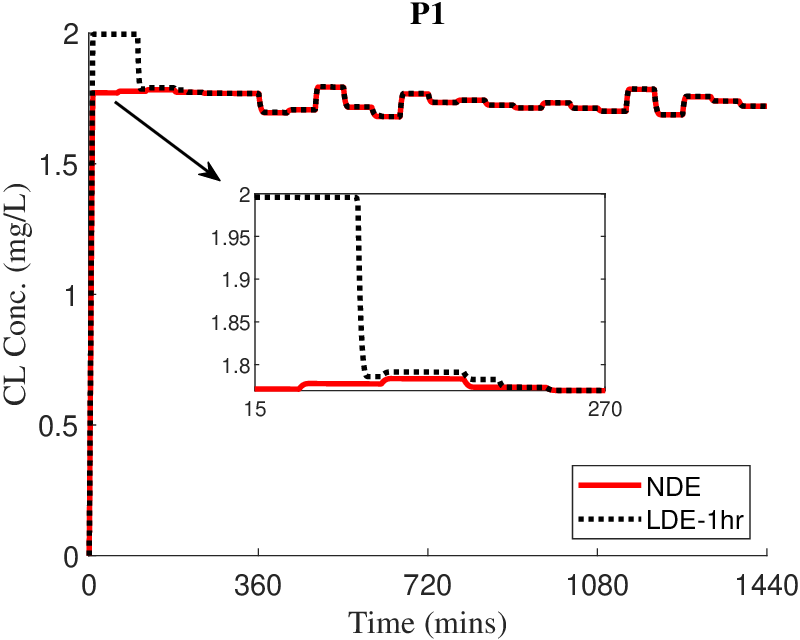}}{}\vspace{-0.15cm} \hspace{0.2cm}
\subfloat[\label{fig:3NLin2b}]{\includegraphics[keepaspectratio=true,scale=0.5]{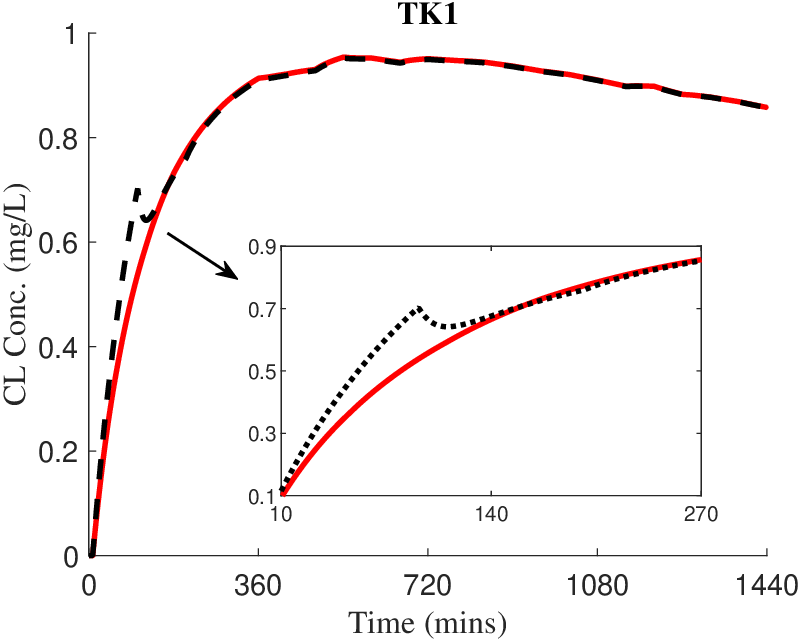}}{}\vspace{-0.15cm} \hspace{0cm}
\subfloat[\label{fig:3NLin2c}]{\includegraphics[keepaspectratio=true,scale=0.5]{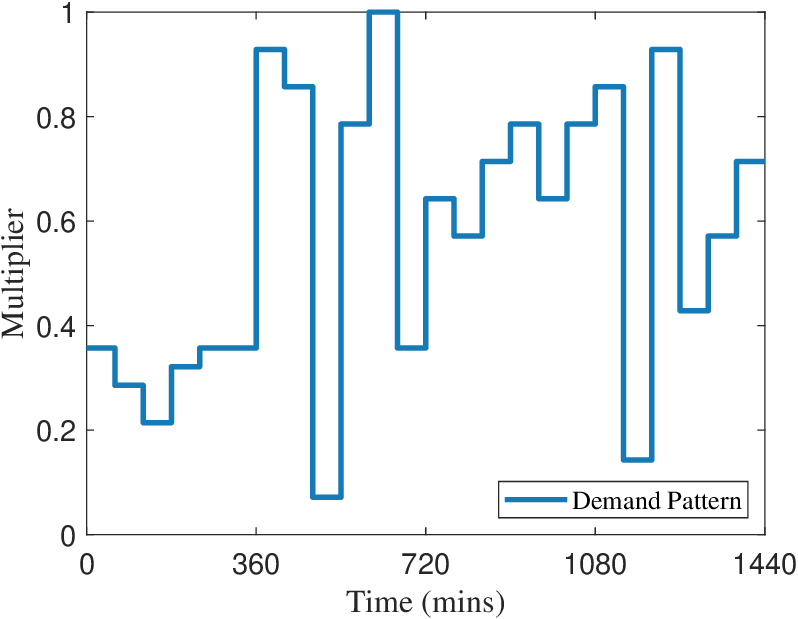}}{}\vspace{-0.15cm} \hspace{0cm}
\caption{Chlorine concentrations at (a) P1 and (b) TK1 of the three-node network with (c) patterned demand at J1. Results are for the nonlinear and linearized models---linearization operating points are updated every 1 hr for all network components.}
\label{fig:3NLin2}
\end{figure} 

The choice of the operating points we linearize around is critical. The narrowest the recurrent window of choosing the operating points, the closest the results to the original model. However, if we choose to update the operating points each water quality time-step then matrices $\breve{\mA}_{11}(t), \breve{\mA}_{12}(t), \breve{\mA}_{21}(t)$ and $\breve{\mA}_{22}(t)$ in Eq. \eqref{systemstatesLDE} should be updated that frequent instead of being updated each hydraulic time-step. Hydraulic time-step is acceptable to be within an hourly scale to reflect the change in demand, while the range for water quality is between minutes and seconds to allow a stable numerical simulation \cite{shangEPANETMSXUserManual2023,seyoum2017integration}. Consequently, updating the aforementioned matrices every WQ time-step adds more computational burden to the simulation which negates the main reason for implementing linearization and model order reduction. On the contrary, widening the window to be more than the hydraulic time-step especially in cases with significant demand change gives inaccurate approximation of the system's behavior. Over and above that, it is important to consider falling within the control algorithm prediction and control horizon to be able to adjust accordingly with the controller input.

With the hydraulic setting of a patterned demand at J1 changing every 1 hr (Fig. \ref{fig:3NLin2c}), the model is linearized around operating points that are taken every 1 hr for each of the network elements. The same sources of chemicals are provided at R1 with zero initial conditions for the other components. Results, shown in Fig. \ref{fig:3NLin2a} and \ref{fig:3NLin2b} for chlorine concentrations at TK1 and P1, exhibit that updating operating points every 1 hr results in accurate representation in comparison to the original model, except for the first hour during which operating points are taken to be the initial concentrations at those elements. To mitigate this issue, operating points are updated after 1-10 minutes from the simulation start. The same approach is followed in scenarios where chemical dosages are increasing locally at some node for elements downstream of this node. 

\begin{figure}[h!]
\centering { \subfloat[\label{fig:MoRRMSEa}]{\includegraphics[keepaspectratio=true,scale=0.5]{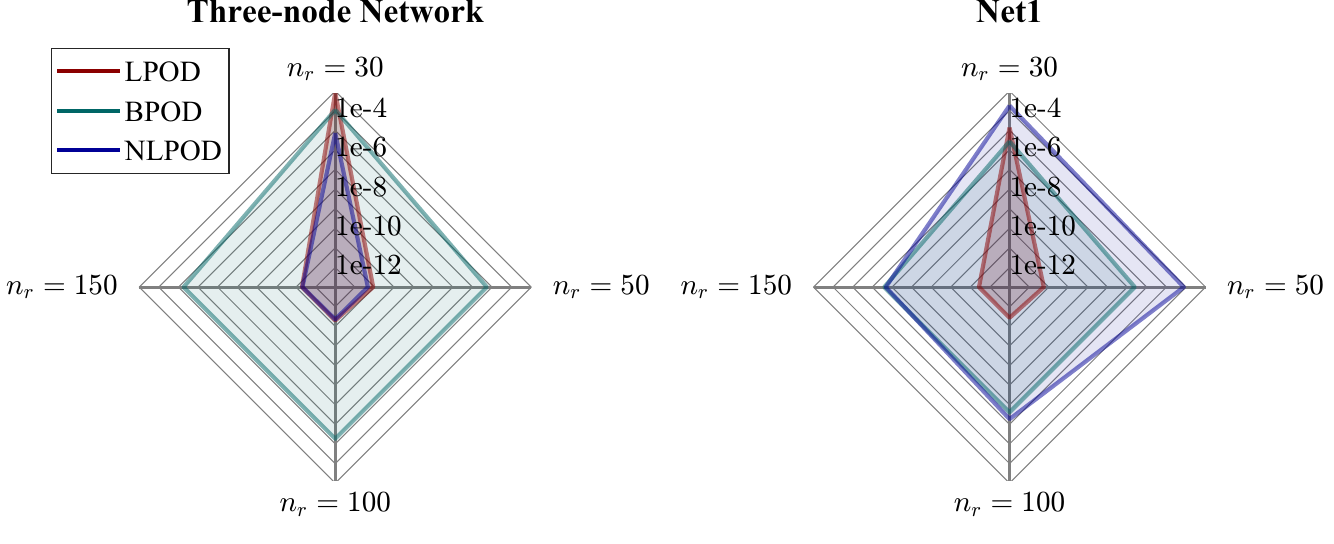}}{}\vspace{-0.2cm} }
{	\subfloat[\label{fig:MoRRMSEb}]{\includegraphics[keepaspectratio=true,scale=0.5]{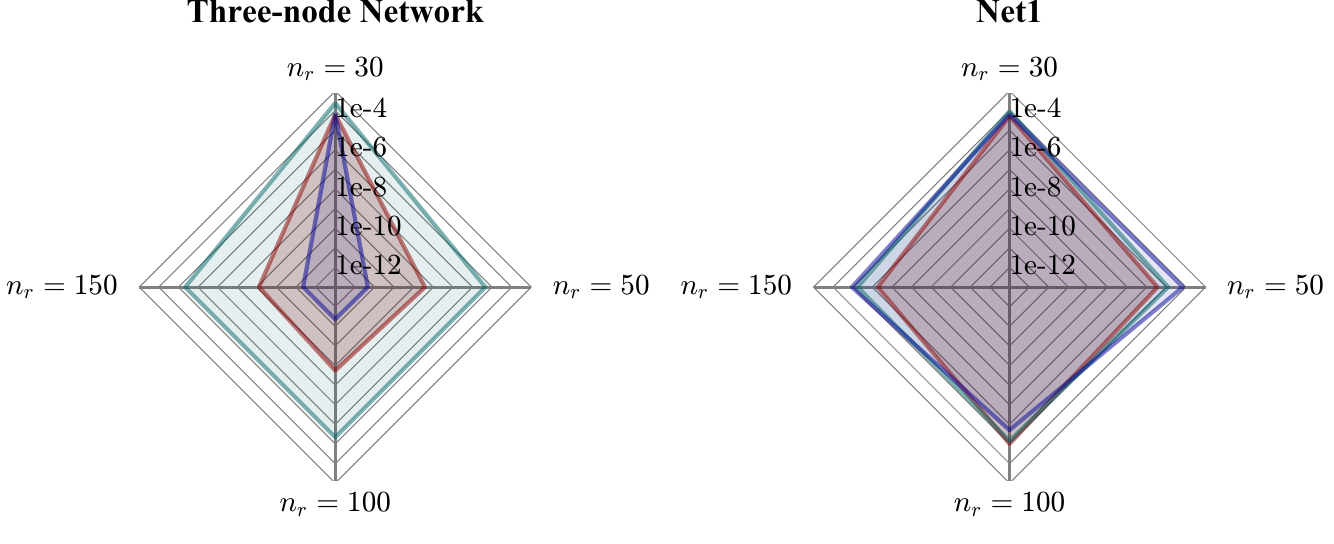}}{}\vspace{-0.2cm}}
\caption{RMSEs for the three proposed MOR methods for the three-node ($n_x=204$) and Net1 ($n_x=482$) networks with (a) zero and (b) non-zero initial conditions for different $n_r$ values.}
\label{fig:MoRRMSE}
\end{figure}

\subsection{MS-WQ Model Order Reduction Performance}
In this section, we assess and compare the performance of each of the proposed model order reduction procedures for multi-species water quality dynamics in terms of accuracy compared to the original full-order model, and computational time. For each network, we apply POD and BPOD on the linearized model and extended POD for the nonlinear model. We refer to these procedures as LPOD, LBPOD, and NLPOD, respectively. We note that we record the computational time needed for assembling the snapshots, obtaining the transformation matrices, and calculating the RMSE between the original and reduced-order model for a specific simulation under the same conditions.   

\begin{figure}[h!]
\centering \vspace{-0.2cm} { \subfloat[\label{fig:TK3N}]{\includegraphics[keepaspectratio=true,scale=0.8]{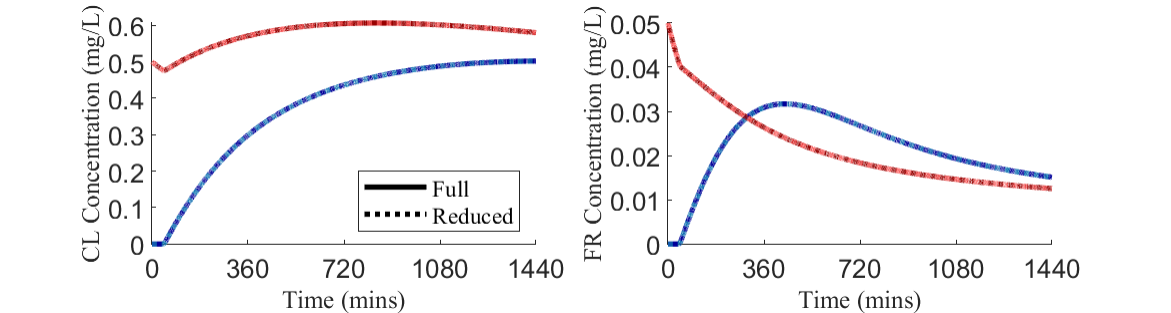}}{}\vspace{-0.1cm} }
{	\subfloat[\label{fig:TKNet1}]{\includegraphics[keepaspectratio=true,scale=0.8]{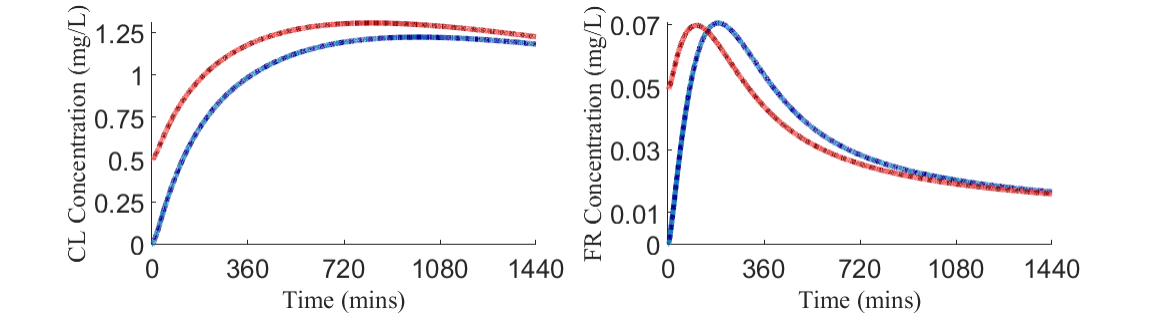}}{}\vspace{-0.2cm}}
\caption{Chlorine and fictitious reactant concentrations evolution at (a) TK1 of three-node network ($n_x=204$) and (b) Tank 11 of Net1 ($n_x=482$) under zero (in blue) and non-zero (in red) initial conditions simulated by full- and reduced-order models with $n_r=30$ for both networks.}
\label{fig:TK3NNet1}
\end{figure}

%\begin{figure}[h!]
%	\hspace{-0.8cm}	{\includegraphics[width=0.58\textwidth]{TK11Net1}}
%	\caption{Chlorine and fictitious reactant concentrations evolution at Tank 11 of Net1 network under zero and non-zero initial conditions simulated by full- and reduced-order models ($n_x=482$ and $n_r=30$).}~\label{fig:TK11Net1}
%\end{figure}

First, we apply the three MOR methods on the three-node and Net1 networks under zero and non-zero initial conditions and static hydraulic profiles. The results shown in Figures \ref{fig:MoRRMSE} and \ref{fig:TK3NNet1} validate that all methods are able to reduce the model dimensions with relatively low RMSEs for different $n_r$ values. These RMSEs get lower with increasing the $n_r$ values and are lower for the scenario of zero initial conditions compared to the case of non-zero initial conditions. For the scenario with non-zero initial conditions, initial chlorine concentrations are 0.5 mg/L network-wide; the initial fictitious reactant concentrations at TK1 in the three-node network and Tank 11 in Net1 are 0.05 mg/L. Fig. \ref{fig:TK3NNet1} shows the chlorine and fictitious reactant concentrations for both scenarios of initial conditions at TK1 of the three-node network and Tank 11 of Net1 for the full-order model and the reduced-order models using all three MOR producers. It is observed that the reduced-order models give almost identical results to the full-order one for the step response at TK1 and for a regular node along the network for Tank 1 for the two scenarios of zero and non-zero initial conditions. On the contrary, these results differ from \cite{wang2022model}, where the POD method was found to have higher errors for the scenario of non-zero initial concentrations under single-species dynamics as the input-output relationship is not correctly captured when the initial values are treated as inputs into the system. In our study, this effect is mitigated by building the offline snapshot with a higher impulse signal by the booster stations which results in favoring the actual locations of booster stations.

Meanwhile, it is worth mentioning that MOR methods' performance is significantly impacted by the locations of the sensors and actuators and their reflection on network-wide observability and controllability. This leads to inaccurate or unstable results in some cases and in some other scenarios. However, the allocation of these sensors and actuators for each network is out of this paper's scope and we solve assuming the predetermination of their locations.  

\subsection{Model Order Reduction Sensitivity to System Hydraulics}

The construction of the transformation matrices $V_r$ and $L_r$ for both methods POD and BPOD is sensitive to the snapshots (i.e., $\mX_m$ and $\mP_m$) constructed offline. These snapshots need to be long enough and representative of the actual reaction between states, inputs, and outputs. That leads to being sensitive to the hydraulic settings of the system while capturing these snapshots and also while applying the desired model reduction. Dynamic hydraulic states in a network reflect the consumers' patterned consumption which can be recorded for a specific network during a specific season \cite{mazzoni2022investigating}.

\begin{figure}[h!]
{\includegraphics[width=\textwidth]{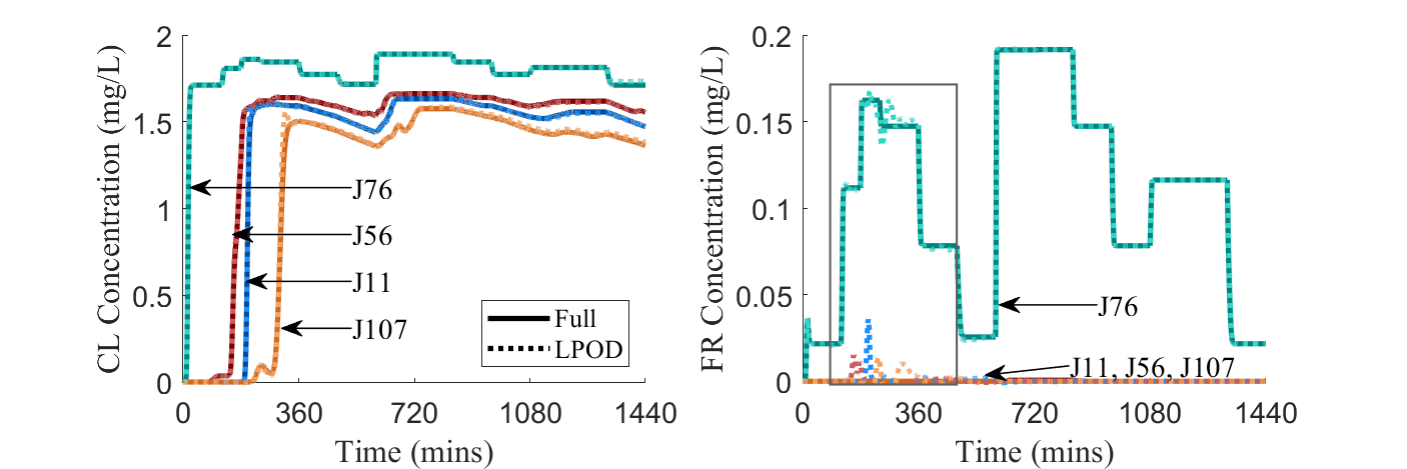}}
\caption{Chlorine and fictitious reactant concentrations evolution at J11, J56, J76, and J107 of FFCL-1 network simulated by full- and reduced-order models. Full-order model results for at each of the junctions are in solid lines, while the LPOD method results are in dashed lines. Number of states for the full-order model is $n_x=10356$ and reduced to $n_r=200$ states.}~\label{fig:FFCLHydro}
\end{figure}

After validating the reliability of the three MOR methods under zero and non-zero initial conditions, we investigate the case of dynamic hydraulic demands for a bigger network; the FFCL-1 network. In Fig. \ref{fig:FFCLHydro}, the evolution of chlorine and the fictitious reactant at J11, J56, J76, and J107 of the FFCL-1 network simulated by full-order model and LPOD-based reduced-order model is presented. Note that, only LPOD is shown, which is representative of the behavior of all other approaches. In this scenario, an input of 0.3 mg/L for the fictitious reactant is inserted at the start of the network (i.e., at the Tank) depicting an early intrusion event. As demonstrated, the LPOD-based ROM is able to trace the concentrations of the chemicals at different junctions, including dead-ends and junctions that are connecting looped pipes. Nonetheless, an oscillatory effect is detected for the fictitious reactant concentrations in the framed zone. This oscillation is formulated as the fictitious reactant being completely consumed by the chlorine at these junctions or at pipes flowing into them (e.g., J76), however, the operating points around which the system is linearized force the fictitious reactant to have false concentrations. Therefore, this effect is illuminated by applying NLPOD and is reduced by updating the operating points more frequently.

Lastly, the computational time recorded for each of the MOR methods implementations on the three tested networks is illustrated in Fig. \ref{fig:CompTime}. For all networks, the NLPOD method requires more computational time as a result of handling the nonlinearity term separately and performing the greedy sampling algorithm. However, the maximum increase in time is around 95 seconds compared to BPOD for the FFCL-1 network, which is considered an acceptable computational time for a network of $n_x=10356$ states.

\begin{figure}[h!]
\centering
\includegraphics[width=0.7\textwidth]{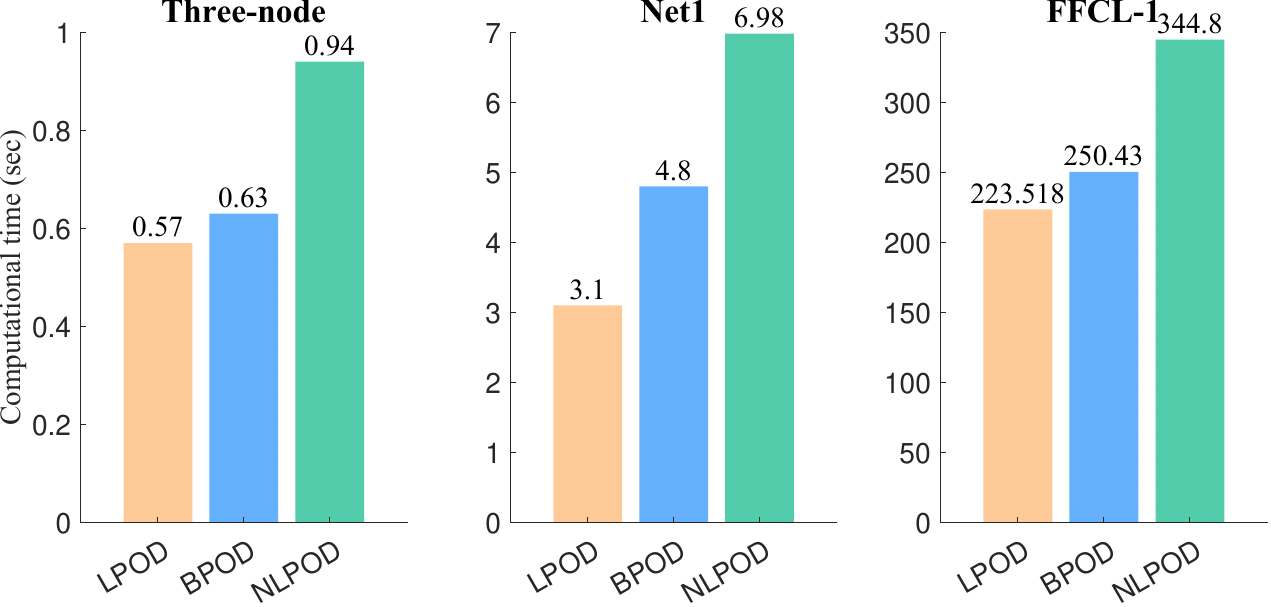}
\caption{Computational time to implement the three MOR methods for three tested networks. Total number of states $n_x$ is 10356 for FFCL-1, 482 for Net1, and 204 for the three-node network. }~\label{fig:CompTime}
\end{figure}

\subsection{Implicit vs. Explicit Discretization Schemes under Control-theoretic Perspective}~\label{sec:ImpExpProf}
As stated in Section "\nameref{sec:TranReactPipes}", the 1-D AR equation can be discretized by implementing either Explicit or Implicit Upwind schemes. The explicit scheme needs to be performed under satisfied CFL condition to ensure stability which requires a small time-step in many cases and hence a higher system dimension. While the implicit scheme is unconditionally stable but requires more complicated mathematical calculations that add to the computational work. Therefore, it has been a pressing question that needs to be answered, "\textit{Which is better: Implicit or Explicit discretization schemes?}". This has been proven to not be an easy question with an easy answer. In our study, we reduce our system's dimensions while applying either of these methods. Nonetheless, while transformation matrices are calculated offline, some system matrices are updated every hydraulic time-step. This adds more computational load with matrices multiplication which is higher with matrix inverse in the case of the implicit scheme. One more point to highlight, although the implicit scheme allows a bigger simulation time-step, a smaller one is more efficient to be able to update the control inputs more frequently. So our question can be formulated as follows, "\textit{From a control-theoretic perspective, which is better: Implicit or Explicit discretization schemes?}"

\begin{figure}[h!]
\centering
\subfloat[\label{fig:ExpImpJ4a}]{\includegraphics[keepaspectratio=true,scale=0.55]{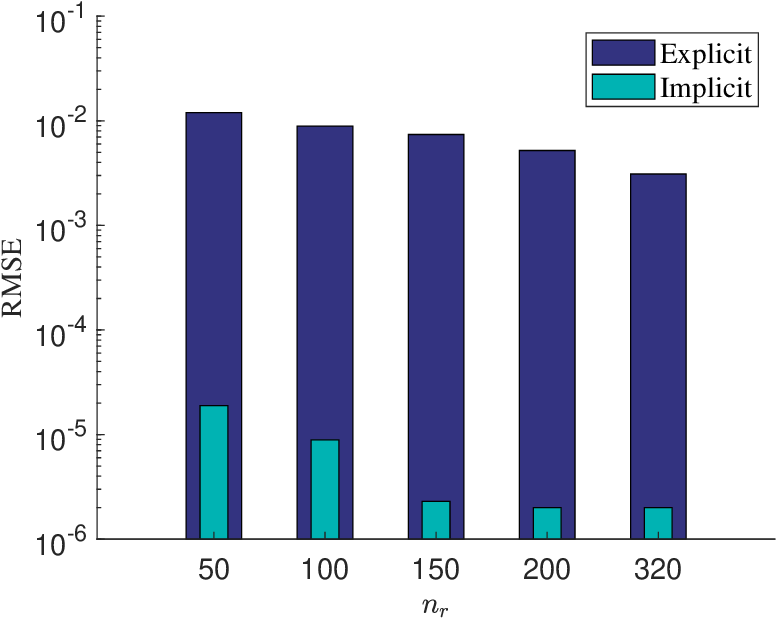}}{}\vspace{-0.05cm} \hspace{-0.01cm}
\subfloat[\label{fig:ExpImpJ4b}]{\includegraphics[keepaspectratio=true,scale=0.55]{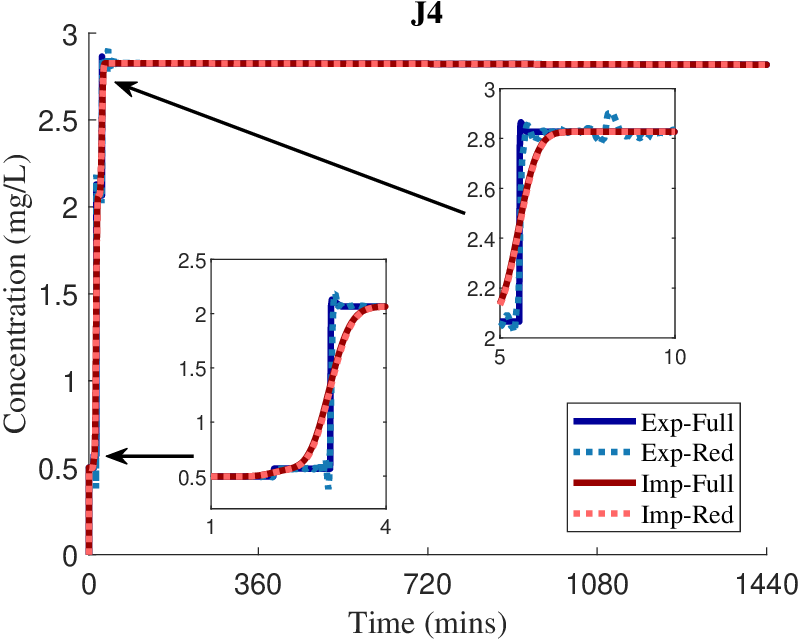}}{}\vspace{-0.05cm} \hspace{-0.3cm}
\caption{(a) RMSEs for the reduced-order Explicit and Implicit Upwind schemes-based models while applying LPOD on Net1 with $n_x=482$ for different $n_r$ values, (b) chlorine concentrations at J4 simulated by the full- and reduced-order models (with $n_r=100$) using both schemes.}
\label{fig:ExpImpJ4}
\end{figure}

As model order reduction is a prior step to applying control to our model, we test both discretization methods' performance while applying the LPOD method for Net1. As demonstrated in Fig. \ref{fig:ExpImpJ4a}, the RMSEs are lower for the implicit than the explicit scheme. In addition, the change in RMSEs by increasing $n_r$ more than 150 is insignificant as the states which get retained do not hold high energy compared to the previously selected ones. On the other hand, the error values for the explicit scheme do not go lower than 0.003 with increasing $n_r$---explained through the following example. Although the CFL condition is satisfied for the explicit scheme, it formulates sharp fronts at points with a relatively significant change in the chemical concentrations as shown in Fig. \ref{fig:ExpImpJ4b}. Fig. \ref{fig:ExpImpJ4b} illustrates chlorine concentrations at J4 for both implicit and explicit schemes and the corresponding reduced models with $n_r=120$ of a full model with $n_x=482$. It is noticeable that the reduced model based on the explicit scheme is consequently affected and shows instability behavior that dampens reaching equilibrium. This performance is recorded under a low Courant number with the network's pipes. To mitigate that, the water quality time-step is required to be reduced to reach higher CN---near but less than 1. Such behavior is avoided when applying the unconditionally stable implicit scheme. 

To that end, the Implicit Upwind scheme gives more accurate results that lead to a more robust control algorithm. The computational burden of this scheme can be lowered using sparse matrix multiplication. The computational time to perform the simulation on Net1 shown in Fig. \ref{fig:ExpImpJ4a} is 32.9 and 43.4 seconds for the Explicit and Implicit Upwind schemes, respectively for the same water quality time-step. However, the implicit scheme retains high accuracy under a higher WQ time-step while requiring lower computational runtime. Therefore, the Implicit Upwind scheme gives more flexibility in choosing the time-step needed to retain real-time control windows while maintaining high accuracy.

\subsection{Real-time Control Implementation of MS-WQM MOR-Based MPC}

The main objective of this paper and the prior investigation of the MOR methods is to integrate them into and apply a real-time control algorithm of chlorine concentrations using the booster stations distributed along the WDN under MS-WQM. We apply the MPC algorithm on the linearized and nonlinearized MS-WQ ROM as explained in Section "\hyperref[sec:MS-MPC]{Real-Time Regulation of MS-WQM via Model Predictive Control and McCormick Relaxations}". As both LPOD and BPOD can reduce the MS-WQM effectively, For the linearized model we apply the BPOD method. On the other hand, we apply the NLPOD for the nonlinear model to obtain the ROM. 

For multi-species water quality control and regulation, while applying the McCormick relaxation the envelopes rely on the limits for both chemicals. For the network's components near the location of the second chemical intrusion, these envelopes put tight boundaries on the chosen value for $z$ by the control problem as the $x_2$ is close to $x_2^{\max}$. On the other hand, for components downstream from this location with lower concentrations for both chemicals the relaxation allows higher and lower values for $z$ which leads to choosing a value of $z$ to be as close to the underestimators so that the control inputs are lower and the cost of chlorine injections is reduced. Additionally, for higher values of the mutual reaction coefficient $k_r$, the effect of relaxation on the chosen control input increases. That is, the proposed relaxed MPC may result in overlooking/underestimating the mutual reaction and, therefore, we lower the upper bound for the fictitious reactant as a procedure integrated into the looped control algorithm repeated each time-step.

As explained in Section "\nameref{sec:CaseStudLin}", it is proposed to update the operating points around which the system is linearized every significant change window (e.g., hydraulic states change). Updating these operating points adds to the computational time by recalculating the matrices, yet it yields a more accurate representation. Therefore, we put a threshold according to which we judge changing these points after applying every control input. 

\begin{figure}[t!]
\centering
\subfloat[\label{fig:3NMPCa}]{\includegraphics[keepaspectratio=true,scale=0.7]{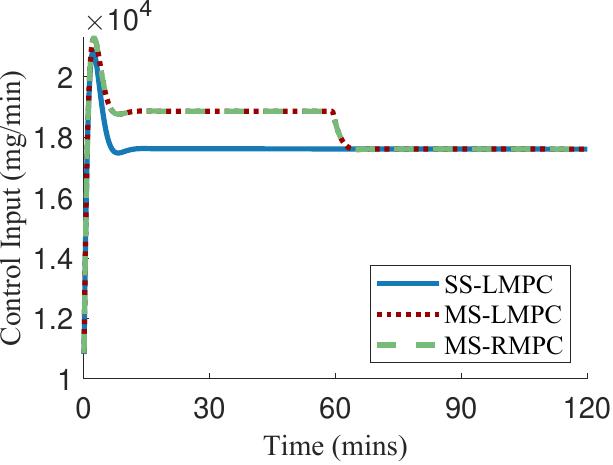}}{}\vspace{-0.05cm} \hspace{0cm}
\subfloat[\label{fig:3NMPCb}]{\includegraphics[keepaspectratio=true,scale=0.7]{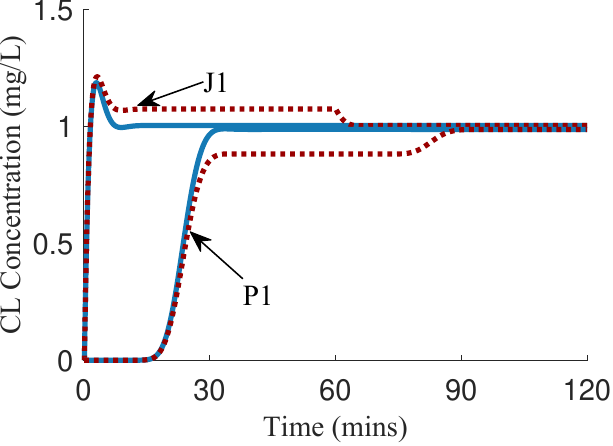}}{}\vspace{-0.05cm} \hspace{-0.1cm}
\caption{(a) Control action $\vu_1$ during 2 hrs of simulation on the three-node network by applying, SS-LMPC: linear single-species-based MPC, MS-LMPC: linearized multi-species-based MPC, and MS-RMPC: nonlinear multi-species-based relaxed MPC, (b) chlorine concentrations at J1 and P1 under another chemical intrusion at J1 for the first hour.}
\label{fig:3NMPC1}
\end{figure} 

By adopting these approaches, we start with applying the MS-WQC MPC-based method on the three-node network under a static hydraulic profile and with a reduced number of states of $n_r=30$ states out of $n_x=204$. The water quality time-step is chosen to be 5 seconds and the control horizon is 10 minutes. The fictitious reactant is discharged into the system at J1 (same location of the booster station) at a concentration of 0.1 mg/L for the first 1 hour of the simulation duration. Fig. \ref{fig:3NMPC1} demonstrates the control actions and the corresponding control response in J1 and P1 under the multi-species linearized ROM, nonlinear ROM, and the single-species ROM that neglects the existence of the other chemical in the system for the first consecutive 2 hours of simulation. For all scenarios chlorine concentrations at J1 and P1 are zeros at the start of the simulation. That is, MPC starts by injecting high chlorine dosage of 21284 mg/min for the case of multi-species dynamics and 20838 mg/min for the single-species model. The control input needed drops to 19158 mg/min and 17596 mg/min for multi-species and single-species dynamics, respectively. After the first hour of simulation, MPC results in the same control actions for both models as the intrusion event is contained. Note that, the second substance's initial concentration for P1 is zero, which leads to the peak control action at the start of the simulation being relatively close as the second substance has not traveled into P1. Comprehensively, this highlights the importance and effectiveness of the adopted MS-WQM and control framework. Furthermore, this difference between the two models' results (i.e., chlorine concentration dynamics and optimal chlorine inputs) increases for more reactive components with chlorine and initial intrusion concentrations which may cause operational issues with limited chlorine availability and/or budget.

Additionally, the linearized MS-WQC problem and relaxed one produce the same performance as illustrated in Fig. \ref{fig:3NMPCa}. We note that for the linearized model, the operating points are updated at the start of the simulation, with applying the peak control action, and by the end of the contamination event. For the relaxed MS-WQC problem, all elements are directly affected by the event resulting in tight envelops and approximating the mutual reaction near its actual value. However, the number of control variables for this procedure is higher for the first hour. To that end, the computational time needed for each of the two control procedures is case-oriented. For this case study, the linearized-based MPC method has computational time of 78 seconds, while it is 93 seconds for the second method.  
\begin{figure}[h!]
\hspace{-0.3cm}
\centering
\subfloat[\label{fig:Net1MPC1}]{\includegraphics[keepaspectratio=true,scale=0.5]{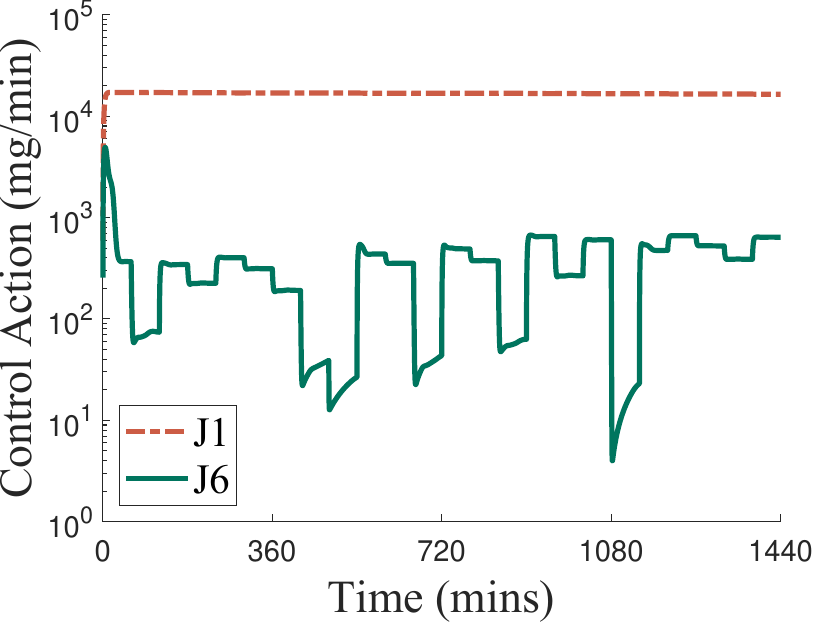}}{}\vspace{-0.05cm} \hspace{0cm}
\subfloat[\label{fig:Net1MPC2}]{\includegraphics[keepaspectratio=true,scale=0.5]{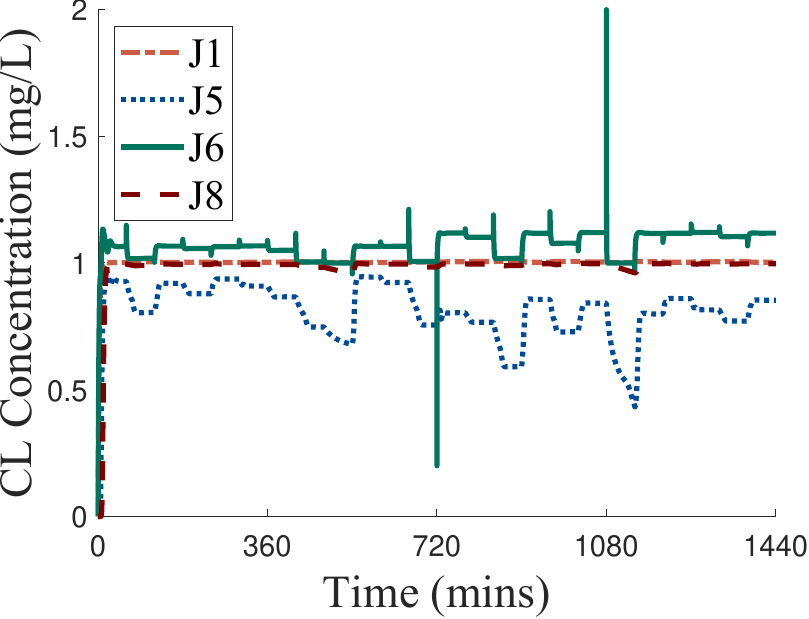}}{}\vspace{-0.05cm} \hspace{0cm}
\subfloat[\label{fig:Net1MPC3}]{\includegraphics[keepaspectratio=true,scale=0.5]{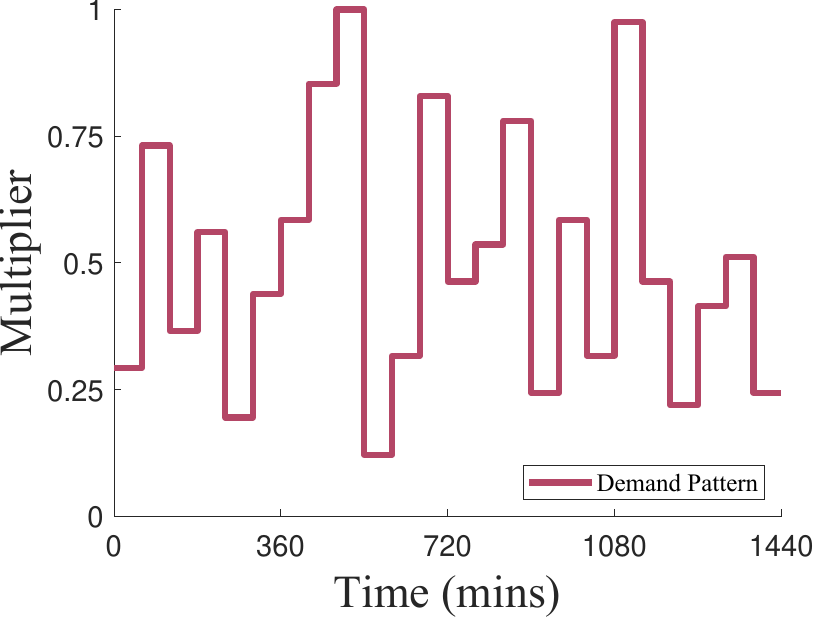}}{}\vspace{-0.05cm} \hspace{-0.1cm}
\caption{(a) MPC control action at Junctions 1 and 6 of Net1, and the corresponding chlorine concentrations at these junctions and Junctions 5 and 8 under (c) patterned demand at Junction 1.}~\label{fig:NetMPC}
\end{figure} 

Next, we apply the proposed MS-WQC approach on Net1 under a dynamic hydraulic profile defined by the patterned demand at Junction 1, Fig. \ref{fig:Net1MPC3}. The FOM has 482 states which are reduced to 50 states. As both control procedures have proved their ability to regulate chlorine concentrations network-wide, we showcase the results from the relaxed MPC procedure only to point out case studies that can take place. In this case study, the water quality time-step is 5 seconds while the control horizon is 10 minutes and the simulation period is 24 hours. The initial concentrations of all chemicals are zeros. The fictitious reactant is set to intrude the system at Junction 6 with a concentration of 0.3 mg/L mid-day. Additionally, chlorine concentrations are limited to 1.2 mg/L for cost reasons. for this case study, we introduce two types of disturbance to the system: a sudden drop in chlorine concentration at Junction 6 to 0.15 mg/L at the 12th hour of the day and a sudden increase to 2 mg/L at the 18th hour. Fig. \ref{fig:Net1MPC1} shows the control action at Junctions 1 and 6, while Fig. \ref{fig:Net1MPC2} demonstrates the corresponding chlorine concentrations at these Junctions and Junctions 5 and 8. For Junction 1, the control action is higher and almost constant at $1.9 \times 10^4$ mg/min as the junction is located at the very start of the network and all the downstream elements are affected by its input. On the other hand, the booster station at Junction 6 acts on the disturbances and the changes at the downstream nodes effectively. Results validate the performance of the control algorithm and how adaptive it is under these disturbances. The run-time recorded for applying the control algorithm for this case study is 278 seconds. 
Likewise, chlorine concentrations are regulated through the FFCL-1 network as Fig. \ref{fig:FFCLMPC1} exhibits. The total number of states of the original model is 10356, while the reduced model has 200 states. Water quality time-step, control horizon, and simulation period are same as the previous case study. Two different fictitious reactants are assumed to be detected, the first one at J76 with initial concentration of 0.3 mg/L while the second one at J89 of 0.2 mg/L. Control actions illustrated in Fig. \ref{fig:FFCLMPCa} are under the condition of hydraulic profile that results in changing flow directions. Yet, the control algorithm recovers effectively and maintain chlorine concentrations within the desired range. In short, the ROMs-based control algorithms guarantee the bounds defined for the inputs and outputs while being tractable for larger networks.

\begin{figure}[t!]
\centering
\subfloat[\label{fig:FFCLMPCa}]{\includegraphics[keepaspectratio=true,scale=0.5]{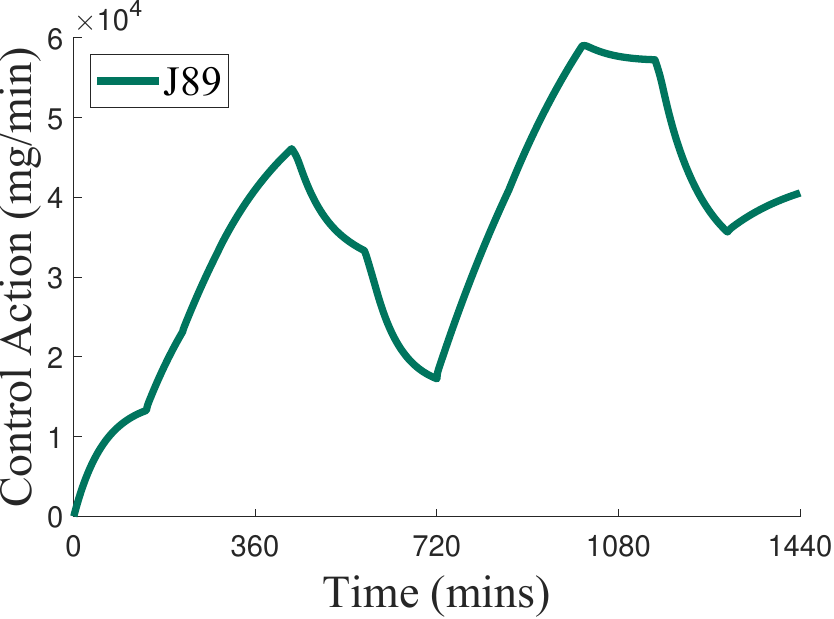}}{}\vspace{-0.05cm} \hspace{0cm}
\subfloat[\label{fig:FFCLMPCb}]{\includegraphics[keepaspectratio=true,scale=0.5]{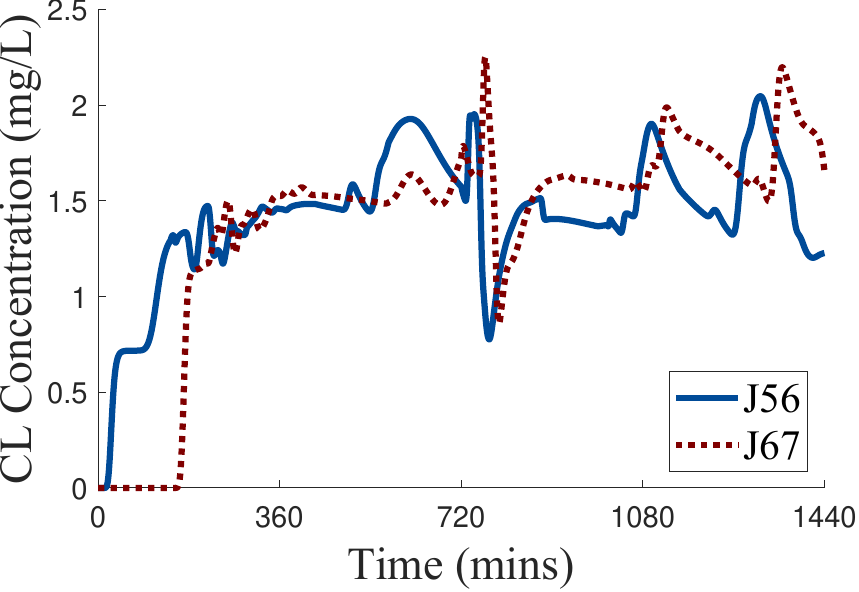}}{}\vspace{-0.05cm} \hspace{-0.1cm}
\caption{(a) Control action at J89 of the FFCL-1 network, (b) the corresponding chlorine concentrations at J56 and J67.}
\label{fig:FFCLMPC1}
\end{figure}

%While applying the McCormick relaxation, the envelopes rely on the limits for both chemicals. For network's components near the location of the second chemical intrusion, these envelopes put tight boundaries on the chosen value for $z$ by the control problem as the $x_2$ is close to $x_2^\max$. 
%%In cases where the second substance has low initial intruded value, these envelopes are tighter compared to events with high initial concentrations resulting in close approximation to the actual system. 
%On another hand, for farther components with lower concentrations for both chemicals the relaxation allows higher and lower values for $z$ which leads to choosing a value of $z$ to be as close to the underestimators so that the control inputs are lower and the cost of chlorine injections is reduced.
%%  in cases where this contaminant is aggressively discharged into the network (e.g., case of leakage of sewage water into WDNs) and the relaxation allows higher and lower values for $z$, the value of $z$ is chosen to be as close to the underestimators so that the control inputs are lower and the cost of chlorine injections is reduced. 
%Additionally, as the mutual reaction coefficient $k_r$ becomes more rapid, the effect of relaxation on the chosen control input increases. That is, the proposed relaxed MPC may result in overlooking/underestimating the mutual reaction and, therefore, we lower the upper bound for the fictitious reactant as a procedure integrated in the looped control algorithm repeated each time-step.

\section{Conclusion, Paper's Limitations, and Recommendations for Future Work}~\label{ConcLimRoc}
Relying on the results from the numerical case studies in Section "\hyperref[sec:CaseStud]{Case Studies}", we answer the research questions posed:
\begin{itemize}[leftmargin=.3in]
\item[$\triangleright$] \textit{A1:} The multi-species water quality model can be effectively linearized around operating points updated every specific moving window according to the hydraulic profile, instantaneous changes, initial conditions, and control actions. However, to achieve the desired accuracy this window is reduced and accordingly, the computational time increases.
\item[$\triangleright$] \textit{A2:} The presented MOR methods yield high accuracy in estimating output concentrations for both chlorine and the fictitious substance in the system. The three MOR procedures: LPOD, BPOD, and NLPOD are able to handle non-zero initial conditions by favoring the control actuators' inputs while building the offline snapshots. Additionally, the NLPOD method requires more computational time to handle and interpolate the nonlinearity in the system, yet, it is still computationally tractable, same for LPOD and BPOD. 
\item[$\triangleright$] \textit{A3:} MPC's behavior depends on the underlying model and its accuracy. Accordingly, the Implicit Upwind scheme is preferred over the Explicit Upwind scheme because of its ability to provide highly-accurate simulation for the full and reduced-order MS-WQM. Moreover, numerical case studies show that the three MOR producers are robust to dynamically changing hydraulic profiles.
\item[$\triangleright$] \textit{A4:} MPC shows robustness and high flexibility in regulating chlorine levels in WDNs under different scenarios of contamination events and hydraulic profiles by applying feedback control on the reduced order model while maintaining affordable computational requirements. Both proposed control procedures, the linearized model- and the relaxed nonlinear model-based show reliable performance while applying adaptive approaches according to the case study considered. These approaches lead to a different level of complexity and computational burden for each of the procedures which results in favoring one procedure over the other according to the case study.
\end{itemize}

Our study is not limitations-free. We highlight these limitations next along with the authors’ future work to be investigated. First, this work used pre-assigned fixed-location booster and sensor locations. Given that these locations impact performance, future work will include optimizing these locations from a control-theoretic perspective.
%First, we apply our framework to networks with pre-distributed booster stations and water quality sensors. The locations of these actuators and sensors affect the performance of the MOR and control of the system. Henceforward, more investigation on their placement from a control-theoretic perspective and accurate model order reduction is recommended for future work. 
Second, additional approaches to model linearization should be explored to potentially exploit offline pre-computed FOM trajectories.
%Second, in the process of linearizing our model the choice of linearization points can be done through offline methods that pre-compute trajectories of the FOM and select linearization regions accordingly; e.g., Trajectory piecewise-linear (TPWL) method \cite{rewienski2003trajectory,white2003trajectory} which is a direction for future inspection. 
Lastly, further work is needed to improve the relaxation method because we expect opportunities to further improve computational performance compared to the linearized model. On the other hand, this study is considered a computational study that is based on a model that has been verified, however, real-time experimental study to verify the considered model and our framework performance under various scenarios is recommended.
%Lastly, the proposed relaxation method (i.e., McCormick envelops) can be tightened in a piecewise way as the tighter the lower and upper bounds, the higher the quality of the relaxation. Accordingly, it can be adopted more effectively under different scenarios of multi-species dynamics in the system while resulting in lower computational time compared to the linearized model.

\subsection{Data Availability Statement}

All data, models, and codes that support the findings of this study are available from the corresponding author upon reasonable request.

\subsection{Acknowledgments}

This work is partially supported by National Science Foundation under grants 1728629, 2015603, 2015671, 2151392, and 2015658.

%\section{Supplemental Materials}
%Sections \comment{S1–S3} are available online in the ASCE Library (\hyperlink{www.ascelibrary.org}{www.ascelibrary.org})

\newpage
\bibliographystyle{IEEEtran}
\bibliography{IEEEabrv,bibfile}

\end{document}